\documentclass[nolinenumbers,twocolumn]{aastex631}

\usepackage{bm}
\usepackage{amsmath}

\newcommand\todo[1]{\textcolor{red}{#1}}
\renewcommand\todo[1]{}

\newcommand{\alteration}[1]{{#1}}

\usepackage{printlen}

\begin{document}

\title{The Case for Edge-On Binaries: An Avenue Toward Comparative Exoplanet Demographics}

\author[0009-0004-9078-5987]{Joseph E. Hand}
\altaffiliation{Dorrit Hoffleit Undergraduate Research Fellow}
\affiliation{Department of Physics and Astronomy, University of Kansas, Lawrence,  KS 66045, USA}
\affiliation{Department of Astronomy, Yale University, New Haven, CT 06511, USA}

\author[0000-0002-4836-1310]{Konstantin Gerbig}
\affiliation{Department of Astronomy, Yale University, New Haven, CT 06511, USA}

\author[0000-0002-7670-670X]{Malena Rice}
\affiliation{Department of Astronomy, Yale University, New Haven, CT 06511, USA}



\begin{abstract}

Most Sun-like and higher-mass stars reside in systems that include one or more gravitationally bound stellar companions. These systems offer an important probe of planet formation in the most common stellar systems, while also providing key insights into how gravitational perturbations and irradiation differences from a companion star alter the outcomes of planet formation. Recent dynamical clues have begun to emerge that reveal systematic, non-random structure in the configurations of many planet-hosting binary systems: in close- to moderate-separation ($s<800$ au) binary star systems, the orbits of exoplanets around individual stellar components are preferentially aligned with the orbital plane of their host stellar binary. In this work, we flip this narrative and search for nearby, edge-on binary star systems that, due to this preferential alignment, are top candidates for radial velocity and transiting exoplanet searches. We present a sample of \alteration{591} moderate-separation, relatively bright \alteration{($G<14$)} \textit{Gaia}-resolved binary star systems in likely near-edge-on configurations. Using a simulated population of exoplanets drawn from transit survey occurrence rate constraints, we provide an overview of the expected planet yields from a targeted search in these systems. We describe the opportunities for comparative exoplanet demographics in the case that \textit{both} stars can be inferred to host edge-on planetary systems -- a configuration toward which the presented sample may be biased, given recent observations of orbit-orbit alignment in exoplanet-hosting binary systems.

\end{abstract}



\section{Introduction}

A sizable fraction of all stars are found in systems with one or more bound stellar companions, such that the role of stellar binarity cannot be ignored when interpreting exoplanet demographics. Companion stars may play an important dynamical role both during and after planet formation, sculpting the exoplanet populations in such systems. Beyond providing a dynamical probe, binary star systems also offer the potential to unveil foundational trends in the outcomes of planet formation, given the opportunity of a ``control sample'' as both stars form in roughly the same conditions and at the same time.

Despite detection biases that disfavor their identification in many large-scale surveys, exoplanet-hosting binary star systems are likely omnipresent across the local solar neighborhood \citep{hirsch2021understanding}. Blending of sources can obfuscate radial velocity (RV) signals while reducing the observed transit depth in photometric surveys, making exoplanets especially difficult to identify in these systems. Nevertheless, hundreds of confirmed and candidate circumstellar exoplanets (on $s-$type orbits) in binary star systems have been identified to date, largely through observations from the space-based \textit{Kepler} \citep{borucki2010kepler} and Transiting Exoplanet Survey Satellite \citep[TESS;][]{ricker2015tess} missions. 

Recent studies of these systems have shown that, considering systems from both \textit{Kepler} \citep{dupuy_orbital_2022} and TESS \citep{christian_possible_2022, Zhang2023dynamical, christian2024wide}, the orbits of transiting exoplanets are preferentially aligned with the orbit of their host stellar binary\alteration{, with typical relative inclinations less than $\sim20-30$\textdegree} -- a configuration that we will call ``orbit-orbit alignment'', following \citet{rice_orbital_2024}. Intriguingly, this trend persists from close-in separations \citep{lester_visual_2023} up to relatively wide sky-projected separations of $s\sim800$ au \citep{christian_possible_2022, gerbig_aligning_2024}, where primordial alignment during star formation should be inefficient \citep[e.g.][]{bate2010chaotic}. The observed alignment potentially reflects dynamical processes, such as dissipative precession during the protoplanetary disk phase \citep{foucart2014evolution, Zanazzi2018torquedamp, gerbig_aligning_2024}, or it may instead arise from primordial sources of alignment \citep[][]{bate2018on}. Notably, the orbit-orbit alignment trend appears to strengthen for decreasing primary-to-companion mass ratio \citep{gerbig_aligning_2024}, as well as relatively small (non-hot Jupiter) short-period exoplanets \citep{christian2024wide}.

In this work, we build upon the observed alignment trend to consider the implications for future surveys. We leverage constraints from the \textit{Gaia} astrometric mission \citep{gaia2016} to identify all bright edge-on binary star systems in the local solar neighborhood. Given that short-period exoplanet orbits for moderate- to wide-separation binaries tend to be well-aligned with the orbit of their host binary, these systems are ideal candidates for exoplanet RV searches, which have maximized signals for near-edge-on orbital configurations. The sample should also include an enhanced rate of transiting exoplanets, with the exact numbers determined by the orbital separation and eccentricity of planets present within the systems, as well as the degree to which they deviate from exactly edge-on. A unique advantage of our approach is that the resulting sample may enable direct comparison of planetary systems within moderately-wide binary systems: if both stars host near-edge-on planetary systems, the planets around each star can be directly compared to provide insights into how environmental factors influence planet formation. 

We note that not all binary exoplanet-hosting systems are expected to exhibit enhanced population-level alignment: in the case that the mutual inclination between orbits is sufficiently large, secular von Zeipel-Lidov-Kozai (ZLK) interactions between the between binary companion and formed planets may further misalign the system \citep{naoz2012on, Naoz2016, Zhang2018}. Hints toward this trend have been observed \citep{rice_orbital_2024}, but primarily in hot-Jupiter-hosting systems that are intrinsically rare. Our focus is on the most common systems -- primarily those with lower-mass exoplanets -- which show a stronger trend toward alignment \citep{christian2024wide} and which lack clear signatures indicative of ZLK oscillations that are seen in the hot Jupiter population.

Our work is outlined as follows. In Section \ref{sec:selection}, we discuss the selection criteria used to construct our sample of edge-on binary systems that exhibit the key properties of promising exoplanet hosts. In Section
\ref{sec:modelling}, we describe the projected yield of exoplanets from a search in these systems. We summarize our findings and further discuss the opportunities presented by this sample in Section \ref{sec:disc}.

\section{Building a Sample: Edge-On Binaries Amenable to RV Follow-Up}\label{sec:selection}

\subsection{Initial sample selection}\label{subsec:init_selection}

\begin{figure}
  \begin{center}
    \includegraphics[width=\linewidth]{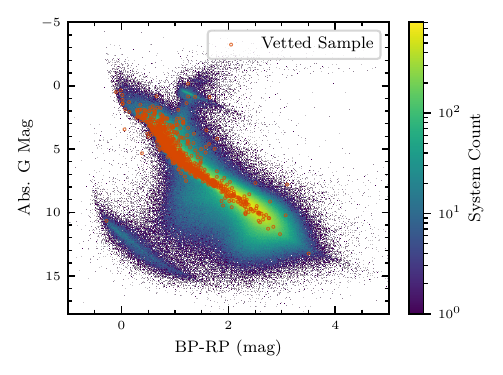}
  \end{center}
  \caption{A color-magnitude diagram of stars in the \citet{el-badry_million_2021} binary catalog, including both the primary and secondary stars in each system. The overlaid red points show stars included in our vetted sample of edge-on binaries.}\label{fig:whole_cmd}
\end{figure}

Our sample was drawn from the \cite{el-badry_million_2021} catalog of 1.8 million binary star systems, identified using photometric and astrometric constraints from the \textit{Gaia} Early Data Release 3 \citep[eDR3;][]{gaia2021edr3}. All binaries in the sample have high-precision astrometric measurements from the \textit{Gaia} mission, and they have been vetted by \citet{el-badry_million_2021} to demonstrate that the binary components have parallaxes consistent with each other and relative proper motions consistent with a Keplerian orbit. A color-magnitude diagram of the full initial sample is shown in Figure \ref{fig:whole_cmd}.

We downselected from this broader set of binaries based on (1) the sky-projected relative velocities of the stars in each system -- specifically, checking whether the binaries are consistent with edge-on orbits; (2) the sky-projected separation between the two stellar components; \alteration{(3) the Renormalized Unit Weight Error (RUWE) to remove poor astrometric solutions and potential unresolved multiple-star sources}; and (4) the magnitudes of the two stellar components, to restrict our sample to stars that are relatively bright and hence more favorable for follow-up observations. We detail each cut in the following paragraphs. 

\begin{figure*} 
\centering
\includegraphics[width=\textwidth]{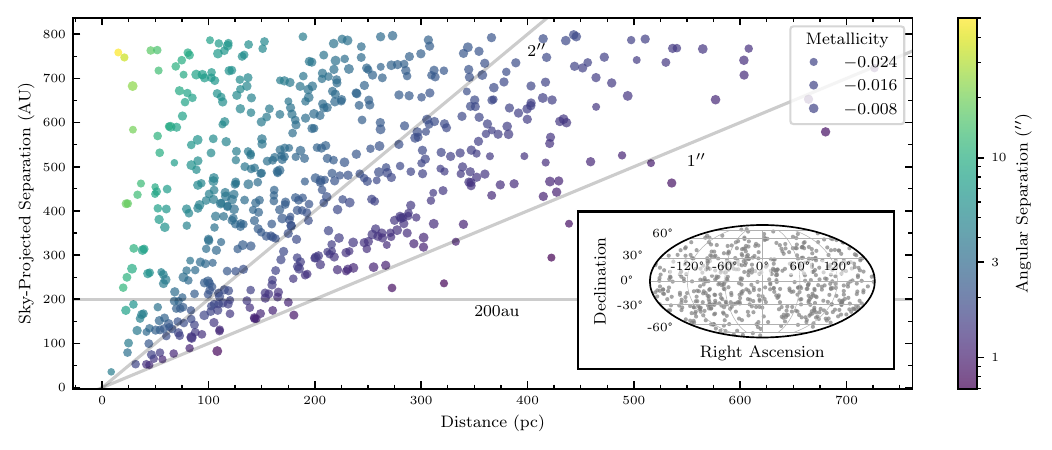}
  \caption{Sky-projected separation of the systems in the final sample, plotted against their
  distance. Also shown is the angular separation of the stars and their metallicity calculated using 
  \texttt{isoclassify}. \alteration{Three} gray reference lines are shown, representing a 200 au sky-projected separation -- indicating the rough (non-sky-projected) binary separation below which exoplanet occurrence is suppressed \citep{moe2021impact} -- a $1\arcsec$ angular separation between the two binary components, which roughly corresponds to the ground-based seeing limit imposed by atmospheric turbulence, \alteration{and a $2\arcsec$ angular separation, providing a practical RV observability reference (since most precision RV spectrographs have fiber or slit sizes up to $\sim1-2\arcsec$). 39\% of the stars in our sample fall between the $1-2\arcsec$ lines, and an additional 2\% fall below the $1\arcsec$ line.} The inset figure in the bottom right shows a Mollweide-projected sky map of the stars in our final sample.}
  \label{fig:sepau}
\end{figure*}

First, we applied a cut to remove binaries that are inconsistent with an edge-on configuration. We used the sky-projected orbit angle, $\gamma$ (adopted from e.g. \citet{tokovinin2015eccentricity}, \citet{behmard2022stellar}, and \citet{rice_orbital_2024}), to carry out this initial vetting step. $\gamma$ is the angle between the two stars' relative position on the sky $\bm{r}$ and their relative proper motion vector $\bm{v}$, i.e.
\begin{align}
    \gamma = \arccos \left(\frac{\bm{r}\cdot \bm{v}}{|\bm{r}||\bm{v}|}\right).
\end{align} In a perfectly edge-on system, $\gamma=0^\circ$ or $\gamma=180^\circ$, depending on the orbital phase at the time of measurement, whereas non-edge-on orientations produce intermediate values of $\gamma$. Thus, $\gamma$ can be used for an initial selection of systems that are consistent with an edge-on orientation while circumventing a full orbit fit. We calculated $\gamma$ for each system using the mean values of their astrometric properties as measured by \textit{Gaia} and drawn from \citet{el-badry_million_2021}. Systems where $\gamma$ was not consistent with either $0^\circ$ or $180^\circ$ within $10^\circ$ were excluded. After removing these, we were left with $189,585$ systems.

Second,  we restricted our sample to binary systems with a sky-projected separation $s<800$ AU. This choice was set following \citet{gerbig_aligning_2024}, to match the empirically-determined separation beyond which a population-level preference for orbit-orbit alignment is no longer confidently observed. Under the assumption that the trend in \citet{christian_possible_2022} and \citet{dupuy_orbital_2022} is sufficiently generalizable, exoplanets in these systems are expected to lie on preferentially edge-on orbits, while those in wider binaries would be better represented by isotropy. After this step, the sample was further reduced to $30,246$ systems.

\alteration{Third, we excluded systems for which either component has RUWE $>1.4$. Such elevated RUWE values indicate a poor astrometric solution, which may be the result of the source itself being a multiple-star system. This cut brought the total number of systems in our sample to $5,187$.}

Lastly, we excluded systems in which  either component has \textit{Gaia} magnitude $G\geq14$. Both stars in each system should therefore be bright and relatively amenable to RV \textbf{or transit} follow-up. This cut brought our list down to \alteration{1,102} systems.

\subsection{Refining the sample with isoclassify and LOFTI}\label{subsec:refining}

We further refined our sample by fitting the orbits for each of the 1,102 systems identified from our initial cuts, further constraining the inclination of each system. We leveraged the \texttt{lofti\_gaia} Python package \citep{pearce2020orbital}, based on the Orbits For The Impatient (OFTI) algorithm presented in \cite{blunt_orbits_2017}, to conduct these orbit fits. \texttt{lofti\_gaia} generates random orbital parameters using the distributions expected under isotropy, then accepts or rejects samples based on a comparison with \textit{Gaia} astrometric constraints for that system, producing a posterior distribution of accepted orbits. To carry out these orbit fits, \texttt{lofti\_gaia} requires input masses for each modeled star. 

We applied the \texttt{isoclassify} Python package \citep{huber2017isoclassify, huber2017asteroseismology, berger2020gaia} to derive stellar masses, radii, and metallicities with associated uncertainties for all stars in our sample. \texttt{isoclassify} uses a grid-based approach to derive stellar parameters by mapping isochrones to a set of input observables. Photometry from the \textit{Gaia} \alteration{$G$ band} and parallax measurements were included as inputs. \alteration{We applied \texttt{isoclassify} with photometric uncertainties inflated to $0.3$ magnitudes to ensure that all solutions converged.}

After deriving stellar masses and associated uncertainties, we ran \texttt{lofti\_gaia} fits with $200$ posterior samples accepted per system, drawing all astrometric parameters directly from the \citet{el-badry_million_2021} binary star catalog. From this, we derived inclinations and uncertainties for each stellar binary system, excluding \alteration{6} that failed to converge (likely indicating poor astrometric solutions). Because the inclination posteriors deviate significantly from Gaussian, we define our reported binary inclinations and uncertainties as the 50th percentile of the distribution ($i$) and the difference between that value and the 68th ($\sigma_{i, \rm{upper}}$) and 16th ($\sigma_{i, \rm{lower}}$) percentile. Of the remaining \alteration{1,096} converged systems, we selected the highest-confidence edge-on systems, with $85^\circ<i<95^\circ$, $\sigma_{i, \rm{upper}}<10^\circ$, and $\sigma_{i, \rm{lower}}<10^\circ$, as our final sample, resulting in \alteration{591} systems. 

\subsection{Final sample overview}\label{sec:characterization}

\begin{figure}
    \includegraphics[width=\linewidth]{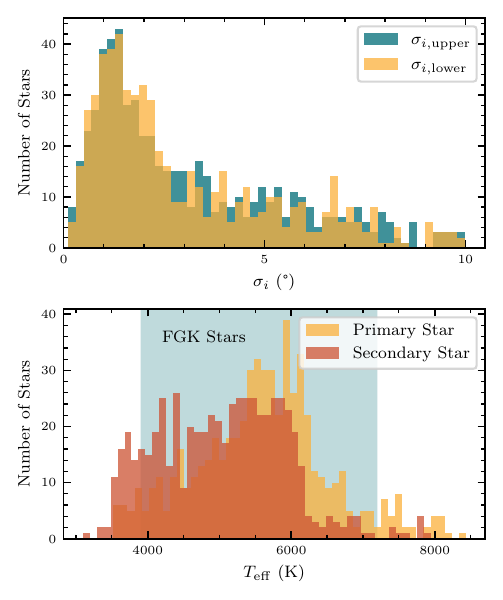}
    \caption{\textit{Top: } Upper and lower asymmetric uncertainties of the system inclinations in our edge-on binaries sample, calculated using \texttt{lofti\_gaia}. \textit{Bottom: } Effective temperatures of the stars, calculated from \textit{Gaia} photometry using \texttt{isoclassify}. The range of temperatures compatible with FGK-type stars is shaded in blue.}
    \label{fig:inc_err_teff}
\end{figure}

Our final sample, with properties listed in Table \ref{tab:binary_systems}, consists of \alteration{591} binary star systems. In each system, both stars have \alteration{$G<14$}, and the components have a sky-projected separation $s<800$ AU. The \alteration{1182} stars in our sample (including both primary and secondary stars), overlaid on the full \citet{el-badry_million_2021} catalog, are shown in Figure~\ref{fig:whole_cmd}. The properties of the sample, including the distribution of systems across the sky, are shown in Figure~\ref{fig:sepau}. The distribution of inclination uncertainties and stellar effective temperatures represented in the sample are provided in Figure \ref{fig:inc_err_teff}. 

\section{Projected Exoplanet Yield}\label{sec:modelling}

To determine the projected exoplanet yield, we modeled the projected radial velocity and transit signals anticipated under the assumption of a \textit{Kepler}-like distribution of exoplanets \citep{kunimoto_searching_2020} (Section \ref{subsec:simulated_sample}). We leveraged this simulated population to demonstrate the recoverability of exoplanets within the sample (Sections \ref{subsec:expected_RV} and \ref{subsec:transit_likelihood}), and we discuss the potential of searches for longer-period exoplanets in Section~\ref{subsec:longer_period}. Finally, we identified and discuss the known exoplanets in the sample (Section \ref{subsec:cross-ref}).

\subsection{Generating a simulated short-period sample}
\label{subsec:simulated_sample}

First, we simulated the projected detection rate for an exoplanet survey across our full sample. Throughout this section, we assume an underlying planet population with properties comparable to those observed in known single-star systems. While exoplanet occurrence rates are lower in close binary systems \citep{hirsch2021understanding}, binary companions with separation $a>200$ au have negligible observed impact on the exoplanet occurrence around each star \citep{moe2021impact}. Our sample includes some binaries with $s<200$ au; however, these are minimum, sky-projected separations and therefore may still represent compelling search targets in many cases. Nevertheless, we include a conservative $s=200$ au line in Figure \ref{fig:sepau} for reference, and we report values considering both our full sample and the $s>200$ au sample.

We consider only short- to medium-period ($P<400\;\text{days}$), small ($<6R_{\oplus}$) exoplanets when determining our projected detection rates. Specifically, our estimates include the parameter space spanned by both the occurrence rate estimates of \citet{kunimoto_searching_2020} and the mass-radius relation from \citet{parviainen_spright_2023}. The restriction to small planets may underestimate the true exoplanet yield: if higher-mass aligned exoplanets also reside in our sample, they would be more easily identified through an RV search. On the other hand, the observed trend toward alignment may be weakened for giant, hot Jupiter exoplanets \citep{christian2024wide}, such that their inclusion within our aligned sample could lead to an overestimated detection rate. Furthermore, short-period giant planets are relatively rare -- found around only $\sim1\%$ of stars \citep{wright2012frequency, fressin2013false, wittenmyer2020cold, beleznay2022exploring} -- such that the majority of detected exoplanets would most likely be smaller. Therefore, we do not expect that the inclusion of these short-period giant planets would significantly change our projected yields. We comment further on wider-orbiting planets in Section \ref{subsec:longer_period}.


For consistency with the \citet{kunimoto_searching_2020} occurrence rates, we consider only main-sequence, FGK stars in our yield analyses. As shown in Figure~\ref{fig:inc_err_teff}, the majority of the stars in our edge-on binaries sample \alteration{($1047/1182$)} lie within the FGK temperature range, with a small tail toward hotter stars. We removed \alteration{$5$} additional potentially post-main-sequence stars through a sample cut to $\log g > 4.0$ dex, though we note that these targets could also be useful to search for exoplanets that are less well-represented by the current census. This leaves \alteration{$1042$} stars remaining. \cite{kunimoto_searching_2020} provides three different distributions of exoplanet occurrence rates for F, G, and K type stars. We accordingly divided our sample into these three categories based on their $T_\mathrm{eff}$, as determined by \texttt{isoclassify}. 

We generated a set of 14, 37, and 57 exoplanets for each F, G, and K star in our sample, respectively. These numbers were chosen to be $30\times$ the average number of exoplanets per star of each classification, as measured in \citet{kunimoto_searching_2020}. Each exoplanet was assigned a radius and period by randomly selecting a radius-period bin from \citet{kunimoto_searching_2020} for the host stellar type, weighted by the bins' mean probability. A point was then drawn at random from this bin according to a logarithmic uniform distribution. For bins with only upper-limit occurrence rates provided, we set the mean to $0$ (corresponding to a 0\% chance of drawing a planet from that bin). 

We then used the \texttt{spright} Python package \citep{parviainen_spright_2023} to assign masses to the planets, leveraging empirical constraints on the mass-radius relationship for small exoplanets ($R<6R_\oplus$). For each planet, three orbital inclinations were adopted using three distinct orbit-orbit alignment schemes: perfect alignment with the host binary's orbit, 20$^{\circ}$ dispersion around perfect alignment (drawing from a Gaussian distribution with $1\sigma=20^{\circ}$, in accordance with the observed trend from \citet{christian_possible_2022}), and random/isotropic alignment, drawn from a \alteration{$\cos i$ distribution} for comparison with field stars. This process resulted in a sample of \alteration{37,334} randomly generated exoplanets matching observed demographics.

\begin{figure*}
  \begin{center}
    \includegraphics[width=\textwidth]{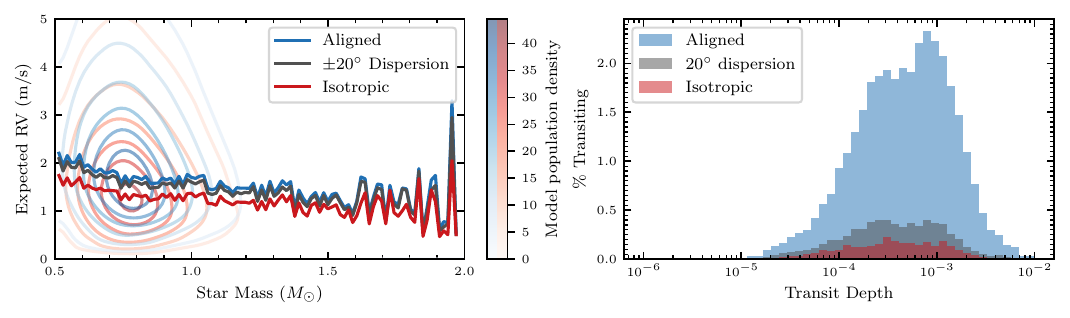}
  \end{center}
  \caption{Simulated exoplanet signals expected for our full edge-on binaries sample. When systems with $s<200$ au are excluded from the sample, the distributions are not substantially affected. \textit{Left:} Semi-amplitudes of the modeled RV signals as a function of host star mass, assuming three different population-level alignment distributions: perfect alignment (\textit{blue}), $20^{\circ}$, dispersion (\textit{grey}), and random/isotropic alignment (\textit{red}). The contours show the distribution of all models, while the dark lines show the mean RV amplitude within each host mass bin. As expected, the RV amplitude decreases with increasing host star mass. There is a factor of $\sim1.6\times$ RV amplitude boost across the full sample between the isotropic alignment scheme, which would be expected for field stars, and the other two alignment schemes, expected for the edge-on binaries in our sample. \textit{Right:} The expected transit signals from the simulated exoplanet population, assuming the same alignment schemes. The aligned exoplanets are significantly more likely to transit than the isotropically oriented ones\alteration{, with the perfectly aligned exoplanets being $11.5$ times as likely to transit, and the exoplanets matching the observed $20^\circ$ dispersion being $2.2$ times as likely to transit.}}\label{fig:rvs}
\end{figure*}

\subsection{Expected short-period planet RV signals}
\label{subsec:expected_RV}

The RV signals from our simulated sample are shown in the left panel of Figure~\ref{fig:rvs}. As anticipated, the aligned population produces systematically higher-amplitude RV signals than an isotropic distribution. 

We predict yields for an RV survey by identifying how many exoplanets in our generated sample meet a set of detection criteria, then dividing by 30 to correct for the inflated number of exoplanets generated (see Section~\ref{subsec:simulated_sample}) and converting to percent yields. Considering the full sample, \alteration{excluding stars outside of the FGK temperature range and stars with $\log g < 4.0$ dex} (\alteration{1,042} stars), \alteration{and loose alignment with $20^\circ$ dispersion}, we predict yields of \alteration{68\%, 4.1\%, and 0.9\%} for companion planets with RV semi-amplitude $K>$ 1, 5, and 10 m/s, respectively. \alteration{By comparison, isotropy would provide predicted yields of 57\%, 3.0\%, and 0.6\%, respectively, for an equivalent sample.} Excluding systems with $s<200$ au (leaving \alteration{940} stars), these estimates were reduced to \alteration{66\%, 4.0\%, and 0.9\% (compared with 55\%, 2.9\%, and 0.5\% assuming isotropy).}

Though many of our simulated signals lie well above the instrumental noise limits of current-generation extreme-precision radial velocity spectrographs ($\sim10-30$ cm/s), a substantive fraction cannot be recovered in practice. One limiting factor is the achievable RV precision: spectrographs typically achieve lower RV precision for stars above the Kraft break \citep[$T_{\rm eff}>6250$ K;][]{kraft1967break} due to Doppler broadening, precluding the detection of most small exoplanets modeled in this work. Considering this limitation, we provide more conservative estimates of exoplanet yields by excluding all stars above the Kraft break from our full sample (leaving \alteration{938} stars), resulting in yields of \alteration{74\%, 4.5\%, and 1.0\%} for $K>$ 1, 5, and 10 m/s, respectively. \alteration{Assuming isotropy, the equivalent predicted yields are 61\%, 3.3\%, and 0.6\%.} Excluding systems with $s<200$ au (leaving \alteration{840} stars) reduces these estimates to \alteration{72\%, 4.4\%, and 1.0\% (compared with 60\%, 3.2\%, and 0.6\% assuming isotropy).}


Another limiting factor is stellar activity, which provides an astrophysical noise source that may mask low-amplitude RV signals. The impact of stellar jitter can be assessed by obtaining vetting spectra prior to initiating a full survey -- for example, via near-infrared spectroscopy using the $\log R_\mathrm{HK}^\prime$ index \citep{noyes_rotation_1984}. Adopting a requirement that $\log R_\mathrm{HK}^\prime<-4.75$ -- indicating relatively inactive stars -- and considering the population-level distribution of $\log R_\mathrm{HK}^\prime$ values for FGK stars identified in \citet{gomes2021stellar}, we anticipate that $\sim$67\% of our identified sample would remain.

\alteration{Lastly, systems with an angular separation smaller than the fiber or slit width of the spectrograph observing them would suffer from contamination of the companion star's flux, which introduces significant challenges in performing RV measurements. For reference, we show $1\arcsec$ and $2\arcsec$ lines on Figure~\ref{fig:sepau}, corresponding to the typical range of fiber/slit sizes for current-era precision RV spectrographs. For systems below these limits, with exact values that will vary by spectrograph, it would be difficult to obtain precise RV measurements using current instrumentation.}

\subsection{Short-period transit likelihood}
\label{subsec:transit_likelihood}

The bias of our sample toward a near-edge-on-configuration also implies that, if the exoplanets are well-aligned with their host stellar binary orbit, they should have a relatively high likelihood of transiting. Therefore, we quantify the transit rate anticipated within our simulated sample, together with the expected transit depths for our exoplanets, with results shown in the right panel of Figure \ref{fig:rvs}. The median expected transit duration for the sample is \alteration{$3.0$} hours, and 90\% of transits have durations between \alteration{$1.0$} and \alteration{$8.3$} hours.

We find that \alteration{6.5}\% of simulated planets transit when assuming a 20$^{\circ}$ dispersion from alignment with their host star, compared with \alteration{3.0}\% transiting for isotropic orientations and \alteration{34.4}\% without dispersion. However, the expected transit depths are relatively shallow -- with a peak likelihood around 500 ppm -- such that they likely would not have been captured by previous transit surveys. For reference, only 478 of the 6156 TOIs on the TESS Project Candidates list (accessed November 15, 2024), excluding false positives, fall at or below this threshold.

Targets within our list are spread across the sky (see inset of Figure \ref{fig:sepau}) -- that is, mostly outside of the \textit{Kepler} field \citep{borucki2010kepler} -- and, while most have been observed by the TESS mission, the angular separation between stars is smaller than the TESS pixel size (21$\arcsec$) in nearly all cases, such that our sample would suffer from severe blending. Therefore, despite the high anticipated transit rate, past surveys likely would not have found most transiting planets within our sample.

We predict the fraction of stars in our sample exhibiting transit signals observable with 1 m ground-based telescopes, with a transit depth threshold of 700 ppm \citep{mallonn_detection_2022}. In the full sample, including stars in systems with $s<200$ au, \alteration{0.9\% of exoplanets modeled} have transit signals observable by ground-based instrumentation, assuming 20$^{\circ}$ dispersion. Excluding \alteration{exoplanets in} systems with $s<200$ au, this estimate is reduced to \alteration{0.8\%}.

\alteration{Beyond individual transiting systems, it is also possible that doubly-transiting binary systems may exist within our sample. Assuming independent transit occurrence rates for each star, the occurrence of such doubly-transiting binary systems would be the square of the occurrence rate of singly-transiting systems. We estimate the occurrence rate of doubly-transiting systems in our sample to be 0.37\%, assuming a $20^\circ$ dispersion in alignment, which would result in an expected 2 such systems within our sample. If strict alignment is assumed, this occurrence rate rises to 11.6\%, yielding an expected 69 such systems in our sample.}

\subsection{Longer-period planets}
\label{subsec:longer_period}
Our simulations include only the signals of short-period, low-mass planets. This is motivated by three primary considerations: (1) the trend toward alignment was found specifically for transiting planets, so that it is unclear whether the trend extends to wider-orbiting planets; (2) based on projected occurrence rates that indicate a paucity of short-period giant planets, small exoplanets would likely constitute the bulk of discoveries at short orbital periods; and (3) short-period planets can be discovered most efficiently, on a timeline of $\lesssim3$ years, through RV surveys. 

If our stellar binaries form via disk fragmentation (see e.g. \citet{offner2023_ppvii}), or if dissipative precession operates to align these systems (see \citet{gerbig_aligning_2024}), however, we may expect that wider-orbiting planets should follow the same orbit-orbit trends observed for the transiting population. As such, a longer-term RV survey for wide companions may also have excellent potential for the discovery of additional planets: long-period giant planets have relatively large RV signals and are found around $\sim30\%$ of FGK stars \citep[e.g.][]{fernandes2019hints}. We also note that our identified population, given its relative proximity, would be a \alteration{potential} target group to search for astrometric planet candidates with datasets such as upcoming \textit{Gaia} releases \citep{espinoza2023prospects, feng2024astrometric}. \alteration{However, the data-quality impact of a close stellar companion will likely complicate the process of astrometric fitting, such that the small astrometric signals from exoplanets may be relatively challenging to detect.}

\subsection{Cross-referencing with known exoplanets/TOIs}
\label{subsec:cross-ref}

To search for known planets in our sample, our full 1,182-star sample was cross-referenced against the NASA Exoplanet Archive ``Planetary Systems'' \citep{PS_Table} table and the TESS Objects of Interest \citep[TOI;][]{guerrero2021tess} catalogue, both accessed on November 15, 2024. With a $20^{\prime\prime}$ search radius around each star, \alteration{we identified three confirmed exoplanets within two systems in the final sample: HD 39855 b, a short-period, non-transiting exoplanet confirmed by \citet{feng_search_2019}; WASP-8 b, a short-period, transiting hot Jupiter confirmed by \citet{queloz_wasp-8b_2010}; and WASP-8 c, a long-period, non transiting exoplanet confirmed by \citet{knutson_friends_2014}.} The absence of additional known transiting exoplanets is unsurprising given the sample size and properties (see Section \ref{subsec:transit_likelihood}). In addition to \alteration{these confirmed exoplanets, four TOIs were identified within the sample: TOI-1717.01, TOI-4175.01, TOI-4646.01, and TOI-6832.01, all with TFOP WG dispositions \citep{akeson_tess_2019} of ``planetary candidate.''}

HD 39855 b has a minimum mass consistent with that of a super-Earth or sub-Neptune \citep[$M_p\sin i=8.5\pm1.5M_{\oplus}$;][]{feng_search_2019}, and our analyses show that the stellar binary has a sky-projected separation $s\sim250$ au with inclination $i=93\substack{+5 \\ -1}^{\circ}$. In the case that HD 39855 b lies on an edge-on orbit that is roughly collinear with its binary host system, the minimum mass would be close to the true mass of the planet. Calculating $\sin i$ for the 200 posterior inclination samples from \texttt{lofti\_gaia} and deriving new posteriors from this distribution, we estimate that $\sin i=0.96\substack{+0.04 \\ -0.15}$ for HD 39855 b under the assumption of orbit-orbit alignment with a 20$^{\circ}$ dispersion, corresponding to a true exoplanet mass $9.3\substack{+2.6 \\ -1.9}M_\oplus$.

\alteration{WASP-8 c has a minimum mass $M_p\sin i=9.45\pm2.26M_\text{Jup}$ \citep{knutson_friends_2014} and is the wider-orbiting of two known exoplanets around WASP-8 A. The other known exoplanet in the system, WASP-8 b, transits and has a well-constrained mass $M_p=2.132 \pm0.080\,M_\text{Jup}$ \citep{bonomo_gaps_2017}. Because WASP-8 c has a long orbital period $P = 4323\pm740\;\text{days}$, its non-transiting configuration is consistent with coplanarity with WASP-8 b, which has inclination $i = 88.55 \pm0.16^\circ$ \citep{bonomo_gaps_2017}. Assuming coplanarity between the two planets, which would also be consistent with edge-on alignment, we estimate the mass of WASP-8 c from 20,000 posterior inclination and $M_p\sin i$ samples, corresponding to a true predicted mass $M_p = 9.43\pm2.27\,M_\text{Jup}$.}

\section{Discussion}\label{sec:disc}

In this work, we identified a sample of \textit{Gaia}-resolved, edge-on binary star systems that, based on the observed orbital geometries of exoplanets in similar systems, are predicted to host an overabundance of edge-on planetary systems (Table \ref{tab:binary_systems}). Our identified systems therefore constitute a promising sample to search for exoplanets through RV or transit observations, which are most sensitive to near-edge-on configurations. We showed that, if these systems are well-aligned with their binary hosts' orbits, a survey of these targets should have a significantly higher yield of exoplanet detections, through both RV and transit searches, than would be expected for a blind search across field stars -- quantified explicitly in Section \ref{sec:modelling}.


If the observed orbit-orbit alignment trend persists, these systems would offer a unique capability to conduct \textit{comparative} demographic studies within binary exoplanet systems. Orbit-orbit aligned systems, with $i\sim90^{\circ}$ for planetary systems around both stars, have the potential to enable direct comparisons of exoplanet populations across pairs of stars that formed under similar initial conditions. Such a comparison would offer fundamental insights into the stochasticity of planet formation, informing whether stars formed in similar environments tend to form similar planets.

Another useful future direction would be the derivation of stellar inclinations for the edge-on binaries sample. Several precisely spin-orbit and orbit-orbit aligned exoplanet-hosting binary systems were previously found in \citet{rice2023orbital} and \citet{rice_orbital_2024}, suggesting the absence of previous dynamical excitation in those systems. Full alignment would be expected if the stellar binary formed via disk fragmentation \citep[see e.g.][]{offner2023_ppvii} or if dissipative precession aligned the system while spin-orbit misalignment was never excited, as explored in \citet{gerbig_aligning_2024}. While the orbital inclinations of non-transiting exoplanet systems cannot be directly inferred, a combination of an edge-on orbital configuration and an edge-on stellar rotation configuration -- indicating two axes of alignment -- would lend confidence to the hypothesis that any identified planets formed quiescently, with an orbit that is also edge-on and that has not undergone significant dynamical excitation to misalign the system \citep[e.g.][]{su2024stellar}.

Stellar inclinations can be derived through a combination of photometric stellar rotation periods and spectroscopic $v\sin i_*$ measurements. While rotation periods can only be inferred for stars with long-lived starspots \citep[e.g.][]{nielsen2013rotation} or asteroseismic signals \citep[e.g.][]{gizon2003determining, kamiaka2018reliability}, a subset of our sample may be suitable for either or both methods. Thus, targeted follow-up observations may help to further constrain the 3D orientations of these systems.


\section*{Acknowledgements}
We thank the anonymous reviewer for their constructive comments on this manuscript. We are grateful to the Dorrit Hoffleit Undergraduate Research Scholarship program at Yale University, which provided support for this project. M.R. acknowledges support from Heising-Simons Foundation Grant \#2023-4478.

This work has made use of data from the European Space Agency (ESA) mission {\it Gaia} (\url{https://www.cosmos.esa.int/gaia}), processed by the {\it Gaia} Data Processing and Analysis Consortium (DPAC, \url{https://www.cosmos.esa.int/web/gaia/dpac/consortium}). Funding for the DPAC has been provided by national institutions, in particular the institutions participating in the {\it Gaia} Multilateral Agreement.

This research has made use of the NASA Exoplanet Archive, which is operated by the California Institute of Technology, under contract with the National Aeronautics and Space Administration under the Exoplanet Exploration Program.

\facilities{Exoplanet Archive, \textit{Gaia}}

\software{\texttt{astropy} \citep{astropy2013, astropy2018, astropy2022}, \texttt{isoclassify} \citep{huber2017isoclassify, huber2017asteroseismology, berger2020gaia}, \texttt{lofti\_gaia} \citep{pearce2020orbital}, \texttt{gnu parallel} \citep{tange_2024_11247979}, \texttt{matplotlib} \citep{hunter2007matplotlib}, \texttt{mwdust} \citep{bovy2016galactic}, \texttt{numpy} \citep{oliphant2006guide, walt2011numpy, harris2020array}, \texttt{pandas} \citep{mckinney2010data}, \texttt{scipy} \citep{virtanen2020scipy}, \texttt{spright} \citep{parviainen_spright_2023}, \texttt{tqdm} \citep{costa-luis_tqdm_2024}}


\bibliography{main}{}

\begin{thebibliography}{}
\expandafter\ifx\csname natexlab\endcsname\relax\def\natexlab#1{#1}\fi
\providecommand{\url}[1]{\href{#1}{#1}}
\providecommand{\dodoi}[1]{doi:~\href{http://doi.org/#1}{\nolinkurl{#1}}}
\providecommand{\doeprint}[1]{\href{http://ascl.net/#1}{\nolinkurl{http://ascl.net/#1}}}
\providecommand{\doarXiv}[1]{\href{https://arxiv.org/abs/#1}{\nolinkurl{https://arxiv.org/abs/#1}}}

\bibitem[{Akeson \& Christiansen(2019)}]{akeson_tess_2019}
Akeson, R., \& Christiansen, J. 2019, 233, 140.09.
\newblock \url{https://ui.adsabs.harvard.edu/abs/2019AAS...23314009A}

\bibitem[{{Astropy Collaboration} {et~al.}(2013){Astropy Collaboration},
  {Robitaille}, {Tollerud}, {Greenfield}, {Droettboom}, {Bray}, {Aldcroft},
  {Davis}, {Ginsburg}, {Price-Whelan}, {Kerzendorf}, {Conley}, {Crighton},
  {Barbary}, {Muna}, {Ferguson}, {Grollier}, {Parikh}, {Nair}, {Unther},
  {Deil}, {Woillez}, {Conseil}, {Kramer}, {Turner}, {Singer}, {Fox}, {Weaver},
  {Zabalza}, {Edwards}, {Azalee Bostroem}, {Burke}, {Casey}, {Crawford},
  {Dencheva}, {Ely}, {Jenness}, {Labrie}, {Lim}, {Pierfederici}, {Pontzen},
  {Ptak}, {Refsdal}, {Servillat}, \& {Streicher}}]{astropy2013}
{Astropy Collaboration}, {Robitaille}, T.~P., {Tollerud}, E.~J., {et~al.} 2013,
  \aap, 558, A33, \dodoi{10.1051/0004-6361/201322068}

\bibitem[{{Astropy Collaboration} {et~al.}(2018){Astropy Collaboration},
  {Price-Whelan}, {Sip{\H{o}}cz}, {G{\"u}nther}, {Lim}, {Crawford}, {Conseil},
  {Shupe}, {Craig}, {Dencheva}, {Ginsburg}, {VanderPlas}, {Bradley},
  {P{\'e}rez-Su{\'a}rez}, {de Val-Borro}, {Aldcroft}, {Cruz}, {Robitaille},
  {Tollerud}, {Ardelean}, {Babej}, {Bach}, {Bachetti}, {Bakanov}, {Bamford},
  {Barentsen}, {Barmby}, {Baumbach}, {Berry}, {Biscani}, {Boquien}, {Bostroem},
  {Bouma}, {Brammer}, {Bray}, {Breytenbach}, {Buddelmeijer}, {Burke},
  {Calderone}, {Cano Rodr{\'\i}guez}, {Cara}, {Cardoso}, {Cheedella}, {Copin},
  {Corrales}, {Crichton}, {D'Avella}, {Deil}, {Depagne}, {Dietrich}, {Donath},
  {Droettboom}, {Earl}, {Erben}, {Fabbro}, {Ferreira}, {Finethy}, {Fox},
  {Garrison}, {Gibbons}, {Goldstein}, {Gommers}, {Greco}, {Greenfield},
  {Groener}, {Grollier}, {Hagen}, {Hirst}, {Homeier}, {Horton}, {Hosseinzadeh},
  {Hu}, {Hunkeler}, {Ivezi{\'c}}, {Jain}, {Jenness}, {Kanarek}, {Kendrew},
  {Kern}, {Kerzendorf}, {Khvalko}, {King}, {Kirkby}, {Kulkarni}, {Kumar},
  {Lee}, {Lenz}, {Littlefair}, {Ma}, {Macleod}, {Mastropietro}, {McCully},
  {Montagnac}, {Morris}, {Mueller}, {Mumford}, {Muna}, {Murphy}, {Nelson},
  {Nguyen}, {Ninan}, {N{\"o}the}, {Ogaz}, {Oh}, {Parejko}, {Parley}, {Pascual},
  {Patil}, {Patil}, {Plunkett}, {Prochaska}, {Rastogi}, {Reddy Janga},
  {Sabater}, {Sakurikar}, {Seifert}, {Sherbert}, {Sherwood-Taylor}, {Shih},
  {Sick}, {Silbiger}, {Singanamalla}, {Singer}, {Sladen}, {Sooley},
  {Sornarajah}, {Streicher}, {Teuben}, {Thomas}, {Tremblay}, {Turner},
  {Terr{\'o}n}, {van Kerkwijk}, {de la Vega}, {Watkins}, {Weaver}, {Whitmore},
  {Woillez}, {Zabalza}, \& {Astropy Contributors}}]{astropy2018}
{Astropy Collaboration}, {Price-Whelan}, A.~M., {Sip{\H{o}}cz}, B.~M., {et~al.}
  2018, \aj, 156, 123, \dodoi{10.3847/1538-3881/aabc4f}

\bibitem[{{Astropy Collaboration} {et~al.}(2022){Astropy Collaboration},
  {Price-Whelan}, {Lim}, {Earl}, {Starkman}, {Bradley}, {Shupe}, {Patil},
  {Corrales}, {Brasseur}, {N{\"o}the}, {Donath}, {Tollerud}, {Morris},
  {Ginsburg}, {Vaher}, {Weaver}, {Tocknell}, {Jamieson}, {van Kerkwijk},
  {Robitaille}, {Merry}, {Bachetti}, {G{\"u}nther}, {Aldcroft},
  {Alvarado-Montes}, {Archibald}, {B{\'o}di}, {Bapat}, {Barentsen},
  {Baz{\'a}n}, {Biswas}, {Boquien}, {Burke}, {Cara}, {Cara}, {Conroy},
  {Conseil}, {Craig}, {Cross}, {Cruz}, {D'Eugenio}, {Dencheva}, {Devillepoix},
  {Dietrich}, {Eigenbrot}, {Erben}, {Ferreira}, {Foreman-Mackey}, {Fox},
  {Freij}, {Garg}, {Geda}, {Glattly}, {Gondhalekar}, {Gordon}, {Grant},
  {Greenfield}, {Groener}, {Guest}, {Gurovich}, {Handberg}, {Hart},
  {Hatfield-Dodds}, {Homeier}, {Hosseinzadeh}, {Jenness}, {Jones}, {Joseph},
  {Kalmbach}, {Karamehmetoglu}, {Ka{\l}uszy{\'n}ski}, {Kelley}, {Kern},
  {Kerzendorf}, {Koch}, {Kulumani}, {Lee}, {Ly}, {Ma}, {MacBride}, {Maljaars},
  {Muna}, {Murphy}, {Norman}, {O'Steen}, {Oman}, {Pacifici}, {Pascual},
  {Pascual-Granado}, {Patil}, {Perren}, {Pickering}, {Rastogi}, {Roulston},
  {Ryan}, {Rykoff}, {Sabater}, {Sakurikar}, {Salgado}, {Sanghi}, {Saunders},
  {Savchenko}, {Schwardt}, {Seifert-Eckert}, {Shih}, {Jain}, {Shukla}, {Sick},
  {Simpson}, {Singanamalla}, {Singer}, {Singhal}, {Sinha}, {Sip{\H{o}}cz},
  {Spitler}, {Stansby}, {Streicher}, {{\v{S}}umak}, {Swinbank}, {Taranu},
  {Tewary}, {Tremblay}, {de Val-Borro}, {Van Kooten}, {Vasovi{\'c}}, {Verma},
  {de Miranda Cardoso}, {Williams}, {Wilson}, {Winkel}, {Wood-Vasey}, {Xue},
  {Yoachim}, {Zhang}, {Zonca}, \& {Astropy Project Contributors}}]{astropy2022}
{Astropy Collaboration}, {Price-Whelan}, A.~M., {Lim}, P.~L., {et~al.} 2022,
  \apj, 935, 167, \dodoi{10.3847/1538-4357/ac7c74}

\bibitem[{{Bate}(2018)}]{bate2018on}
{Bate}, M.~R. 2018, \mnras, 475, 5618, \dodoi{10.1093/mnras/sty169}

\bibitem[{{Bate} {et~al.}(2010){Bate}, {Lodato}, \&
  {Pringle}}]{bate2010chaotic}
{Bate}, M.~R., {Lodato}, G., \& {Pringle}, J.~E. 2010, \mnras, 401, 1505,
  \dodoi{10.1111/j.1365-2966.2009.15773.x}

\bibitem[{Behmard {et~al.}(2022)Behmard, Dai, \& Howard}]{behmard2022stellar}
Behmard, A., Dai, F., \& Howard, A.~W. 2022, \aj, 163, 160

\bibitem[{{Beleznay} \& {Kunimoto}(2022)}]{beleznay2022exploring}
{Beleznay}, M., \& {Kunimoto}, M. 2022, \mnras, 516, 75,
  \dodoi{10.1093/mnras/stac2179}

\bibitem[{Berger {et~al.}(2020)Berger, Huber, Van~Saders, Gaidos, Tayar, \&
  Kraus}]{berger2020gaia}
Berger, T.~A., Huber, D., Van~Saders, J.~L., {et~al.} 2020, \aj, 159, 280

\bibitem[{Blunt {et~al.}(2017)Blunt, Nielsen, De~Rosa, Konopacky, Ryan, Wang,
  Pueyo, Rameau, Marois, Marchis, Macintosh, Graham, Duchene, \&
  Schneider}]{blunt_orbits_2017}
Blunt, S., Nielsen, E.~L., De~Rosa, R.~J., {et~al.} 2017, The Astronomical
  Journal, 153, 229, \dodoi{10.3847/1538-3881/aa6930}

\bibitem[{Bonomo {et~al.}(2017)Bonomo, Desidera, Benatti, Borsa, Crespi,
  Damasso, Lanza, Sozzetti, Lodato, Marzari, Boccato, Claudi, Cosentino,
  Covino, Gratton, Maggio, Micela, Molinari, Pagano, Piotto, Poretti,
  Smareglia, Affer, Biazzo, Bignamini, Esposito, Giacobbe, Hébrard, Malavolta,
  Maldonado, Mancini, Martinez~Fiorenzano, Masiero, Nascimbeni, Pedani, Rainer,
  \& Scandariato}]{bonomo_gaps_2017}
Bonomo, A.~S., Desidera, S., Benatti, S., {et~al.} 2017, Astronomy and
  Astrophysics, 602, A107, \dodoi{10.1051/0004-6361/201629882}

\bibitem[{{Borucki} {et~al.}(2010){Borucki}, {Koch}, {Basri}, {Batalha},
  {Brown}, {Caldwell}, {Caldwell}, {Christensen-Dalsgaard}, {Cochran},
  {DeVore}, {Dunham}, {Dupree}, {Gautier}, {Geary}, {Gilliland}, {Gould},
  {Howell}, {Jenkins}, {Kondo}, {Latham}, {Marcy}, {Meibom}, {Kjeldsen},
  {Lissauer}, {Monet}, {Morrison}, {Sasselov}, {Tarter}, {Boss}, {Brownlee},
  {Owen}, {Buzasi}, {Charbonneau}, {Doyle}, {Fortney}, {Ford}, {Holman},
  {Seager}, {Steffen}, {Welsh}, {Rowe}, {Anderson}, {Buchhave}, {Ciardi},
  {Walkowicz}, {Sherry}, {Horch}, {Isaacson}, {Everett}, {Fischer}, {Torres},
  {Johnson}, {Endl}, {MacQueen}, {Bryson}, {Dotson}, {Haas}, {Kolodziejczak},
  {Van Cleve}, {Chandrasekaran}, {Twicken}, {Quintana}, {Clarke}, {Allen},
  {Li}, {Wu}, {Tenenbaum}, {Verner}, {Bruhweiler}, {Barnes}, \&
  {Prsa}}]{borucki2010kepler}
{Borucki}, W.~J., {Koch}, D., {Basri}, G., {et~al.} 2010, Science, 327, 977,
  \dodoi{10.1126/science.1185402}

\bibitem[{Bovy {et~al.}(2016)Bovy, Rix, Green, Schlafly, \&
  Finkbeiner}]{bovy2016galactic}
Bovy, J., Rix, H.-W., Green, G.~M., Schlafly, E.~F., \& Finkbeiner, D.~P. 2016,
  \apj, 818, 130

\bibitem[{Christian {et~al.}(2022)Christian, Vanderburg, Becker, Yahalomi,
  Pearce, Zhou, Collins, Kraus, Stassun, de~Beurs, Ricker, Vanderspek, Latham,
  Winn, Seager, Jenkins, Abe, Agabi, Amado, Baker, Barkaoui, Benkhaldoun,
  Benni, Berberian, Berlind, Bieryla, Esparza-Borges, Bowen, Brown, Buchhave,
  Burke, Buttu, Cadieux, Caldwell, Charbonneau, Chazov, Chimaladinne, Collins,
  Combs, Conti, Crouzet, de~Leon, Deljookorani, Diamond, Doyon, Dragomir,
  Dransfield, Essack, Evans, Fukui, Gan, Esquerdo, Gillon, Girardin, Guerra,
  Guillot, Habich, Henriksen, Hoch, Isogai, Jehin, Jensen, Johnson, Livingston,
  Kielkopf, Kim, Kawauchi, Krushinsky, Kunzle, Laloum, Leger, Lewin, Mallia,
  Massey, Mori, McLeod, Mékarnia, Mireles, Mishevskiy, Tamura, Murgas, Narita,
  Naves, Nelson, Osborn, Palle, Parviainen, Plavchan, Pozuelos, Rabus, Relles,
  López, Quinn, Schmider, Schlieder, Schwarz, Shporer, Sibbald, Srdoc,
  Stibbards, Stickler, Suarez, Stockdale, Tan, Terada, Triaud, Tronsgaard,
  Waalkes, Wang, Watanabe, Wenceslas, Wingham, Wittrock, \&
  Ziegler}]{christian_possible_2022}
Christian, S., Vanderburg, A., Becker, J., {et~al.} 2022, The Astronomical
  Journal, 163, 207, \dodoi{10.3847/1538-3881/ac517f}

\bibitem[{{Christian} {et~al.}(2024){Christian}, {Vanderburg}, {Becker},
  {Kraus}, {Pearce}, {Collins}, {Rice}, {Jensen}, {Baker}, {Benni}, {Bieryla},
  {Binnenfeld}, {Collins}, {Conti}, {Evans}, {Girardin}, {Gregorio}, {Mazeh},
  {Murgas}, {Panahi}, {Pozuelos}, {Relles}, {Rodriguez Frustaglia}, {Schwarz},
  {Srdoc}, {Stockdale}, {Tan}, {Waalkes}, {Wang}, {Wittrock}, \&
  {Zucker}}]{christian2024wide}
{Christian}, S., {Vanderburg}, A., {Becker}, J., {et~al.} 2024, arXiv e-prints,
  arXiv:2405.10379, \dodoi{10.48550/arXiv.2405.10379}

\bibitem[{Costa-Luis {et~al.}(2024)Costa-Luis, Larroque, Altendorf, Mary,
  richardsheridan, Korobov, Yorav-Raphael, Ivanov, Bargull, Rodrigues, Chen,
  Dektyarev, mjstevens777, Pagel, Zugnoni, JC, CrazyPython, Newey, Lee,
  pgajdos, Todd, Malmgren, redbug312, Desh, Nechaev, Górny, Boyle, Nordlund,
  MapleCCC, \& McCracken}]{costa-luis_tqdm_2024}
Costa-Luis, C.~d., Larroque, S.~K., Altendorf, K., {et~al.} 2024, tqdm: {A}
  fast, {Extensible} {Progress} {Bar} for {Python} and {CLI},  Zenodo,
  \dodoi{10.5281/zenodo.13207611}

\bibitem[{Dupuy {et~al.}(2022)Dupuy, Kraus, Kratter, Rizzuto, Mann, Huber, \&
  Ireland}]{dupuy_orbital_2022}
Dupuy, T.~J., Kraus, A.~L., Kratter, K.~M., {et~al.} 2022, Monthly Notices of
  the Royal Astronomical Society, 512, 648, \dodoi{10.1093/mnras/stac306}

\bibitem[{El-Badry {et~al.}(2021)El-Badry, Rix, \&
  Heintz}]{el-badry_million_2021}
El-Badry, K., Rix, H.-W., \& Heintz, T.~M. 2021, Monthly Notices of the Royal
  Astronomical Society, 506, 2269, \dodoi{10.1093/mnras/stab323}

\bibitem[{{Espinoza-Retamal} {et~al.}(2023){Espinoza-Retamal}, {Zhu}, \&
  {Petrovich}}]{espinoza2023prospects}
{Espinoza-Retamal}, J.~I., {Zhu}, W., \& {Petrovich}, C. 2023, \aj, 166, 231,
  \dodoi{10.3847/1538-3881/ad00b9}

\bibitem[{{Feng}(2024)}]{feng2024astrometric}
{Feng}, F. 2024, arXiv e-prints, arXiv:2403.08226,
  \dodoi{10.48550/arXiv.2403.08226}

\bibitem[{Feng {et~al.}(2019)Feng, Crane, Xuesong~Wang, Teske, Shectman, Díaz,
  Thompson, Jones, \& Butler}]{feng_search_2019}
Feng, F., Crane, J.~D., Xuesong~Wang, S., {et~al.} 2019, The Astrophysical
  Journal Supplement Series, 242, 25, \dodoi{10.3847/1538-4365/ab1b16}

\bibitem[{{Fernandes} {et~al.}(2019){Fernandes}, {Mulders}, {Pascucci},
  {Mordasini}, \& {Emsenhuber}}]{fernandes2019hints}
{Fernandes}, R.~B., {Mulders}, G.~D., {Pascucci}, I., {Mordasini}, C., \&
  {Emsenhuber}, A. 2019, \apj, 874, 81, \dodoi{10.3847/1538-4357/ab0300}

\bibitem[{{Foucart} \& {Lai}(2014)}]{foucart2014evolution}
{Foucart}, F., \& {Lai}, D. 2014, \mnras, 445, 1731,
  \dodoi{10.1093/mnras/stu1869}

\bibitem[{{Fressin} {et~al.}(2013){Fressin}, {Torres}, {Charbonneau}, {Bryson},
  {Christiansen}, {Dressing}, {Jenkins}, {Walkowicz}, \&
  {Batalha}}]{fressin2013false}
{Fressin}, F., {Torres}, G., {Charbonneau}, D., {et~al.} 2013, \apj, 766, 81,
  \dodoi{10.1088/0004-637X/766/2/81}

\bibitem[{{Gaia Collaboration} {et~al.}(2016){Gaia Collaboration}, {Prusti},
  {de Bruijne}, {Brown}, {Vallenari}, {Babusiaux}, {Bailer-Jones}, {Bastian},
  {Biermann}, {Evans}, {Eyer}, {Jansen}, {Jordi}, {Klioner}, {Lammers},
  {Lindegren}, {Luri}, {Mignard}, {Milligan}, {Panem}, {Poinsignon},
  {Pourbaix}, {Randich}, {Sarri}, {Sartoretti}, {Siddiqui}, {Soubiran},
  {Valette}, {van Leeuwen}, {Walton}, {Aerts}, {Arenou}, {Cropper}, {Drimmel},
  {H{\o}g}, {Katz}, {Lattanzi}, {O'Mullane}, {Grebel}, {Holland}, {Huc},
  {Passot}, {Bramante}, {Cacciari}, {Casta{\~n}eda}, {Chaoul}, {Cheek}, {De
  Angeli}, {Fabricius}, {Guerra}, {Hern{\'a}ndez}, {Jean-Antoine-Piccolo},
  {Masana}, {Messineo}, {Mowlavi}, {Nienartowicz}, {Ord{\'o}{\~n}ez-Blanco},
  {Panuzzo}, {Portell}, {Richards}, {Riello}, {Seabroke}, {Tanga},
  {Th{\'e}venin}, {Torra}, {Els}, {Gracia-Abril}, {Comoretto},
  {Garcia-Reinaldos}, {Lock}, {Mercier}, {Altmann}, {Andrae}, {Astraatmadja},
  {Bellas-Velidis}, {Benson}, {Berthier}, {Blomme}, {Busso}, {Carry},
  {Cellino}, {Clementini}, {Cowell}, {Creevey}, {Cuypers}, {Davidson}, {De
  Ridder}, {de Torres}, {Delchambre}, {Dell'Oro}, {Ducourant}, {Fr{\'e}mat},
  {Garc{\'\i}a-Torres}, {Gosset}, {Halbwachs}, {Hambly}, {Harrison}, {Hauser},
  {Hestroffer}, {Hodgkin}, {Huckle}, {Hutton}, {Jasniewicz}, {Jordan},
  {Kontizas}, {Korn}, {Lanzafame}, {Manteiga}, {Moitinho}, {Muinonen},
  {Osinde}, {Pancino}, {Pauwels}, {Petit}, {Recio-Blanco}, {Robin}, {Sarro},
  {Siopis}, {Smith}, {Smith}, {Sozzetti}, {Thuillot}, {van Reeven}, {Viala},
  {Abbas}, {Abreu Aramburu}, {Accart}, {Aguado}, {Allan}, {Allasia},
  {Altavilla}, {{\'A}lvarez}, {Alves}, {Anderson}, {Andrei}, {Anglada Varela},
  {Antiche}, {Antoja}, {Ant{\'o}n}, {Arcay}, {Atzei}, {Ayache}, {Bach},
  {Baker}, {Balaguer-N{\'u}{\~n}ez}, {Barache}, {Barata}, {Barbier}, {Barblan},
  {Baroni}, {Barrado y Navascu{\'e}s}, {Barros}, {Barstow}, {Becciani},
  {Bellazzini}, {Bellei}, {Bello Garc{\'\i}a}, {Belokurov}, {Bendjoya},
  {Berihuete}, {Bianchi}, {Bienaym{\'e}}, {Billebaud}, {Blagorodnova},
  {Blanco-Cuaresma}, {Boch}, {Bombrun}, {Borrachero}, {Bouquillon}, {Bourda},
  {Bouy}, {Bragaglia}, {Breddels}, {Brouillet}, {Br{\"u}semeister},
  {Bucciarelli}, {Budnik}, {Burgess}, {Burgon}, {Burlacu}, {Busonero}, {Buzzi},
  {Caffau}, {Cambras}, {Campbell}, {Cancelliere}, {Cantat-Gaudin}, {Carlucci},
  {Carrasco}, {Castellani}, {Charlot}, {Charnas}, {Charvet}, {Chassat},
  {Chiavassa}, {Clotet}, {Cocozza}, {Collins}, {Collins}, {Costigan}, {Crifo},
  {Cross}, {Crosta}, {Crowley}, {Dafonte}, {Damerdji}, {Dapergolas}, {David},
  {David}, {De Cat}, {de Felice}, {de Laverny}, {De Luise}, {De March}, {de
  Martino}, {de Souza}, {Debosscher}, {del Pozo}, {Delbo}, {Delgado},
  {Delgado}, {di Marco}, {Di Matteo}, {Diakite}, {Distefano}, {Dolding}, {Dos
  Anjos}, {Drazinos}, {Dur{\'a}n}, {Dzigan}, {Ecale}, {Edvardsson}, {Enke},
  {Erdmann}, {Escolar}, {Espina}, {Evans}, {Eynard Bontemps}, {Fabre},
  {Fabrizio}, {Faigler}, {Falc{\~a}o}, {Farr{\`a}s Casas}, {Faye}, {Federici},
  {Fedorets}, {Fern{\'a}ndez-Hern{\'a}ndez}, {Fernique}, {Fienga}, {Figueras},
  {Filippi}, {Findeisen}, {Fonti}, {Fouesneau}, {Fraile}, {Fraser}, {Fuchs},
  {Furnell}, {Gai}, {Galleti}, {Galluccio}, {Garabato}, {Garc{\'\i}a-Sedano},
  {Gar{\'e}}, {Garofalo}, {Garralda}, {Gavras}, {Gerssen}, {Geyer}, {Gilmore},
  {Girona}, {Giuffrida}, {Gomes}, {Gonz{\'a}lez-Marcos},
  {Gonz{\'a}lez-N{\'u}{\~n}ez}, {Gonz{\'a}lez-Vidal}, {Granvik}, {Guerrier},
  {Guillout}, {Guiraud}, {G{\'u}rpide}, {Guti{\'e}rrez-S{\'a}nchez}, {Guy},
  {Haigron}, {Hatzidimitriou}, {Haywood}, {Heiter}, {Helmi}, {Hobbs},
  {Hofmann}, {Holl}, {Holland}, {Hunt}, {Hypki}, {Icardi}, {Irwin}, {Jevardat
  de Fombelle}, {Jofr{\'e}}, {Jonker}, {Jorissen}, {Julbe}, {Karampelas},
  {Kochoska}, {Kohley}, {Kolenberg}, {Kontizas}, {Koposov}, {Kordopatis},
  {Koubsky}, {Kowalczyk}, {Krone-Martins}, {Kudryashova}, {Kull}, {Bachchan},
  {Lacoste-Seris}, {Lanza}, {Lavigne}, {Le Poncin-Lafitte}, {Lebreton},
  {Lebzelter}, {Leccia}, {Leclerc}, {Lecoeur-Taibi}, {Lemaitre}, {Lenhardt},
  {Leroux}, {Liao}, {Licata}, {Lindstr{\o}m}, {Lister}, {Livanou}, {Lobel},
  {L{\"o}ffler}, {L{\'o}pez}, {Lopez-Lozano}, {Lorenz}, {Loureiro},
  {MacDonald}, {Magalh{\~a}es Fernandes}, {Managau}, {Mann}, {Mantelet},
  {Marchal}, {Marchant}, {Marconi}, {Marie}, {Marinoni}, {Marrese},
  {Marschalk{\'o}}, {Marshall}, {Mart{\'\i}n-Fleitas}, {Martino}, {Mary},
  {Matijevi{\v{c}}}, {Mazeh}, {McMillan}, {Messina}, {Mestre}, {Michalik},
  {Millar}, {Miranda}, {Molina}, {Molinaro}, {Molinaro}, {Moln{\'a}r},
  {Moniez}, {Montegriffo}, {Monteiro}, {Mor}, {Mora}, {Morbidelli}, {Morel},
  {Morgenthaler}, {Morley}, {Morris}, {Mulone}, {Muraveva}, {Musella},
  {Narbonne}, {Nelemans}, {Nicastro}, {Noval}, {Ord{\'e}novic},
  {Ordieres-Mer{\'e}}, {Osborne}, {Pagani}, {Pagano}, {Pailler}, {Palacin},
  {Palaversa}, {Parsons}, {Paulsen}, {Pecoraro}, {Pedrosa}, {Pentik{\"a}inen},
  {Pereira}, {Pichon}, {Piersimoni}, {Pineau}, {Plachy}, {Plum}, {Poujoulet},
  {Pr{\v{s}}a}, {Pulone}, {Ragaini}, {Rago}, {Rambaux}, {Ramos-Lerate},
  {Ranalli}, {Rauw}, {Read}, {Regibo}, {Renk}, {Reyl{\'e}}, {Ribeiro},
  {Rimoldini}, {Ripepi}, {Riva}, {Rixon}, {Roelens}, {Romero-G{\'o}mez},
  {Rowell}, {Royer}, {Rudolph}, {Ruiz-Dern}, {Sadowski}, {Sagrist{\`a}
  Sell{\'e}s}, {Sahlmann}, {Salgado}, {Salguero}, {Sarasso}, {Savietto},
  {Schnorhk}, {Schultheis}, {Sciacca}, {Segol}, {Segovia}, {Segransan},
  {Serpell}, {Shih}, {Smareglia}, {Smart}, {Smith}, {Solano}, {Solitro},
  {Sordo}, {Soria Nieto}, {Souchay}, {Spagna}, {Spoto}, {Stampa}, {Steele},
  {Steidelm{\"u}ller}, {Stephenson}, {Stoev}, {Suess}, {S{\"u}veges}, {Surdej},
  {Szabados}, {Szegedi-Elek}, {Tapiador}, {Taris}, {Tauran}, {Taylor},
  {Teixeira}, {Terrett}, {Tingley}, {Trager}, {Turon}, {Ulla}, {Utrilla},
  {Valentini}, {van Elteren}, {Van Hemelryck}, {van Leeuwen}, {Varadi},
  {Vecchiato}, {Veljanoski}, {Via}, {Vicente}, {Vogt}, {Voss}, {Votruba},
  {Voutsinas}, {Walmsley}, {Weiler}, {Weingrill}, {Werner}, {Wevers},
  {Whitehead}, {Wyrzykowski}, {Yoldas}, {{\v{Z}}erjal}, {Zucker}, {Zurbach},
  {Zwitter}, {Alecu}, {Allen}, {Allende Prieto}, {Amorim},
  {Anglada-Escud{\'e}}, {Arsenijevic}, {Azaz}, {Balm}, {Beck}, {Bernstein},
  {Bigot}, {Bijaoui}, {Blasco}, {Bonfigli}, {Bono}, {Boudreault}, {Bressan},
  {Brown}, {Brunet}, {Bunclark}, {Buonanno}, {Butkevich}, {Carret}, {Carrion},
  {Chemin}, {Ch{\'e}reau}, {Corcione}, {Darmigny}, {de Boer}, {de Teodoro}, {de
  Zeeuw}, {Delle Luche}, {Domingues}, {Dubath}, {Fodor}, {Fr{\'e}zouls},
  {Fries}, {Fustes}, {Fyfe}, {Gallardo}, {Gallegos}, {Gardiol}, {Gebran},
  {Gomboc}, {G{\'o}mez}, {Grux}, {Gueguen}, {Heyrovsky}, {Hoar}, {Iannicola},
  {Isasi Parache}, {Janotto}, {Joliet}, {Jonckheere}, {Keil}, {Kim},
  {Klagyivik}, {Klar}, {Knude}, {Kochukhov}, {Kolka}, {Kos}, {Kutka}, {Lainey},
  {LeBouquin}, {Liu}, {Loreggia}, {Makarov}, {Marseille}, {Martayan},
  {Martinez-Rubi}, {Massart}, {Meynadier}, {Mignot}, {Munari}, {Nguyen},
  {Nordlander}, {Ocvirk}, {O'Flaherty}, {Olias Sanz}, {Ortiz}, {Osorio},
  {Oszkiewicz}, {Ouzounis}, {Palmer}, {Park}, {Pasquato}, {Peltzer}, {Peralta},
  {P{\'e}turaud}, {Pieniluoma}, {Pigozzi}, {Poels}, {Prat}, {Prod'homme},
  {Raison}, {Rebordao}, {Risquez}, {Rocca-Volmerange}, {Rosen}, {Ruiz-Fuertes},
  {Russo}, {Sembay}, {Serraller Vizcaino}, {Short}, {Siebert}, {Silva},
  {Sinachopoulos}, {Slezak}, {Soffel}, {Sosnowska}, {Strai{\v{z}}ys}, {ter
  Linden}, {Terrell}, {Theil}, {Tiede}, {Troisi}, {Tsalmantza}, {Tur},
  {Vaccari}, {Vachier}, {Valles}, {Van Hamme}, {Veltz}, {Virtanen}, {Wallut},
  {Wichmann}, {Wilkinson}, {Ziaeepour}, \& {Zschocke}}]{gaia2016}
{Gaia Collaboration}, {Prusti}, T., {de Bruijne}, J.~H.~J., {et~al.} 2016,
  \aap, 595, A1, \dodoi{10.1051/0004-6361/201629272}

\bibitem[{{Gaia Collaboration} {et~al.}(2021){Gaia Collaboration}, {Brown},
  {Vallenari}, {Prusti}, {de Bruijne}, {Babusiaux}, {Biermann}, {Creevey},
  {Evans}, {Eyer}, {Hutton}, {Jansen}, {Jordi}, {Klioner}, {Lammers},
  {Lindegren}, {Luri}, {Mignard}, {Panem}, {Pourbaix}, {Randich}, {Sartoretti},
  {Soubiran}, {Walton}, {Arenou}, {Bailer-Jones}, {Bastian}, {Cropper},
  {Drimmel}, {Katz}, {Lattanzi}, {van Leeuwen}, {Bakker}, {Cacciari},
  {Casta{\~n}eda}, {De Angeli}, {Ducourant}, {Fabricius}, {Fouesneau},
  {Fr{\'e}mat}, {Guerra}, {Guerrier}, {Guiraud}, {Jean-Antoine Piccolo},
  {Masana}, {Messineo}, {Mowlavi}, {Nicolas}, {Nienartowicz}, {Pailler},
  {Panuzzo}, {Riclet}, {Roux}, {Seabroke}, {Sordo}, {Tanga}, {Th{\'e}venin},
  {Gracia-Abril}, {Portell}, {Teyssier}, {Altmann}, {Andrae}, {Bellas-Velidis},
  {Benson}, {Berthier}, {Blomme}, {Brugaletta}, {Burgess}, {Busso}, {Carry},
  {Cellino}, {Cheek}, {Clementini}, {Damerdji}, {Davidson}, {Delchambre},
  {Dell'Oro}, {Fern{\'a}ndez-Hern{\'a}ndez}, {Galluccio}, {Garc{\'\i}a-Lario},
  {Garcia-Reinaldos}, {Gonz{\'a}lez-N{\'u}{\~n}ez}, {Gosset}, {Haigron},
  {Halbwachs}, {Hambly}, {Harrison}, {Hatzidimitriou}, {Heiter},
  {Hern{\'a}ndez}, {Hestroffer}, {Hodgkin}, {Holl}, {Jan{\ss}en}, {Jevardat de
  Fombelle}, {Jordan}, {Krone-Martins}, {Lanzafame}, {L{\"o}ffler}, {Lorca},
  {Manteiga}, {Marchal}, {Marrese}, {Moitinho}, {Mora}, {Muinonen}, {Osborne},
  {Pancino}, {Pauwels}, {Petit}, {Recio-Blanco}, {Richards}, {Riello},
  {Rimoldini}, {Robin}, {Roegiers}, {Rybizki}, {Sarro}, {Siopis}, {Smith},
  {Sozzetti}, {Ulla}, {Utrilla}, {van Leeuwen}, {van Reeven}, {Abbas}, {Abreu
  Aramburu}, {Accart}, {Aerts}, {Aguado}, {Ajaj}, {Altavilla}, {{\'A}lvarez},
  {{\'A}lvarez Cid-Fuentes}, {Alves}, {Anderson}, {Anglada Varela}, {Antoja},
  {Audard}, {Baines}, {Baker}, {Balaguer-N{\'u}{\~n}ez}, {Balbinot}, {Balog},
  {Barache}, {Barbato}, {Barros}, {Barstow}, {Bartolom{\'e}}, {Bassilana},
  {Bauchet}, {Baudesson-Stella}, {Becciani}, {Bellazzini}, {Bernet}, {Bertone},
  {Bianchi}, {Blanco-Cuaresma}, {Boch}, {Bombrun}, {Bossini}, {Bouquillon},
  {Bragaglia}, {Bramante}, {Breedt}, {Bressan}, {Brouillet}, {Bucciarelli},
  {Burlacu}, {Busonero}, {Butkevich}, {Buzzi}, {Caffau}, {Cancelliere},
  {C{\'a}novas}, {Cantat-Gaudin}, {Carballo}, {Carlucci}, {Carnerero},
  {Carrasco}, {Casamiquela}, {Castellani}, {Castro-Ginard}, {Castro Sampol},
  {Chaoul}, {Charlot}, {Chemin}, {Chiavassa}, {Cioni}, {Comoretto}, {Cooper},
  {Cornez}, {Cowell}, {Crifo}, {Crosta}, {Crowley}, {Dafonte}, {Dapergolas},
  {David}, {David}, {de Laverny}, {De Luise}, {De March}, {De Ridder}, {de
  Souza}, {de Teodoro}, {de Torres}, {del Peloso}, {del Pozo}, {Delbo},
  {Delgado}, {Delgado}, {Delisle}, {Di Matteo}, {Diakite}, {Diener},
  {Distefano}, {Dolding}, {Eappachen}, {Edvardsson}, {Enke}, {Esquej}, {Fabre},
  {Fabrizio}, {Faigler}, {Fedorets}, {Fernique}, {Fienga}, {Figueras},
  {Fouron}, {Fragkoudi}, {Fraile}, {Franke}, {Gai}, {Garabato},
  {Garcia-Gutierrez}, {Garc{\'\i}a-Torres}, {Garofalo}, {Gavras}, {Gerlach},
  {Geyer}, {Giacobbe}, {Gilmore}, {Girona}, {Giuffrida}, {Gomel}, {Gomez},
  {Gonzalez-Santamaria}, {Gonz{\'a}lez-Vidal}, {Granvik},
  {Guti{\'e}rrez-S{\'a}nchez}, {Guy}, {Hauser}, {Haywood}, {Helmi}, {Hidalgo},
  {Hilger}, {H{\l}adczuk}, {Hobbs}, {Holland}, {Huckle}, {Jasniewicz},
  {Jonker}, {Juaristi Campillo}, {Julbe}, {Karbevska}, {Kervella}, {Khanna},
  {Kochoska}, {Kontizas}, {Kordopatis}, {Korn}, {Kostrzewa-Rutkowska},
  {Kruszy{\'n}ska}, {Lambert}, {Lanza}, {Lasne}, {Le Campion}, {Le Fustec},
  {Lebreton}, {Lebzelter}, {Leccia}, {Leclerc}, {Lecoeur-Taibi}, {Liao},
  {Licata}, {Lindstr{\o}m}, {Lister}, {Livanou}, {Lobel}, {Madrero Pardo},
  {Managau}, {Mann}, {Marchant}, {Marconi}, {Marcos Santos}, {Marinoni},
  {Marocco}, {Marshall}, {Martin Polo}, {Mart{\'\i}n-Fleitas}, {Masip},
  {Massari}, {Mastrobuono-Battisti}, {Mazeh}, {McMillan}, {Messina},
  {Michalik}, {Millar}, {Mints}, {Molina}, {Molinaro}, {Moln{\'a}r},
  {Montegriffo}, {Mor}, {Morbidelli}, {Morel}, {Morris}, {Mulone}, {Munoz},
  {Muraveva}, {Murphy}, {Musella}, {Noval}, {Ord{\'e}novic}, {Orr{\`u}},
  {Osinde}, {Pagani}, {Pagano}, {Palaversa}, {Palicio}, {Panahi}, {Pawlak},
  {Pe{\~n}alosa Esteller}, {Penttil{\"a}}, {Piersimoni}, {Pineau}, {Plachy},
  {Plum}, {Poggio}, {Poretti}, {Poujoulet}, {Pr{\v{s}}a}, {Pulone}, {Racero},
  {Ragaini}, {Rainer}, {Raiteri}, {Rambaux}, {Ramos}, {Ramos-Lerate}, {Re
  Fiorentin}, {Regibo}, {Reyl{\'e}}, {Ripepi}, {Riva}, {Rixon}, {Robichon},
  {Robin}, {Roelens}, {Rohrbasser}, {Romero-G{\'o}mez}, {Rowell}, {Royer},
  {Rybicki}, {Sadowski}, {Sagrist{\`a} Sell{\'e}s}, {Sahlmann}, {Salgado},
  {Salguero}, {Samaras}, {Sanchez Gimenez}, {Sanna}, {Santove{\~n}a},
  {Sarasso}, {Schultheis}, {Sciacca}, {Segol}, {Segovia}, {S{\'e}gransan},
  {Semeux}, {Shahaf}, {Siddiqui}, {Siebert}, {Siltala}, {Slezak}, {Smart},
  {Solano}, {Solitro}, {Souami}, {Souchay}, {Spagna}, {Spoto}, {Steele},
  {Steidelm{\"u}ller}, {Stephenson}, {S{\"u}veges}, {Szabados}, {Szegedi-Elek},
  {Taris}, {Tauran}, {Taylor}, {Teixeira}, {Thuillot}, {Tonello}, {Torra},
  {Torra}, {Turon}, {Unger}, {Vaillant}, {van Dillen}, {Vanel}, {Vecchiato},
  {Viala}, {Vicente}, {Voutsinas}, {Weiler}, {Wevers}, {Wyrzykowski}, {Yoldas},
  {Yvard}, {Zhao}, {Zorec}, {Zucker}, {Zurbach}, \& {Zwitter}}]{gaia2021edr3}
{Gaia Collaboration}, {Brown}, A.~G.~A., {Vallenari}, A., {et~al.} 2021, \aap,
  649, A1, \dodoi{10.1051/0004-6361/202039657}

\bibitem[{{Gerbig} {et~al.}(2024){Gerbig}, {Rice}, {Zanazzi}, {Christian}, \&
  {Vanderburg}}]{gerbig_aligning_2024}
{Gerbig}, K., {Rice}, M., {Zanazzi}, J.~J., {Christian}, S., \& {Vanderburg},
  A. 2024, \apj, 972, 161, \dodoi{10.3847/1538-4357/ad5f2b}

\bibitem[{{Gizon} \& {Solanki}(2003)}]{gizon2003determining}
{Gizon}, L., \& {Solanki}, S.~K. 2003, \apj, 589, 1009, \dodoi{10.1086/374715}

\bibitem[{{Gomes da Silva} {et~al.}(2021){Gomes da Silva}, {Santos},
  {Adibekyan}, {Sousa}, {Campante}, {Figueira}, {Bossini}, {Delgado-Mena},
  {Monteiro}, {de Laverny}, {Recio-Blanco}, \& {Lovis}}]{gomes2021stellar}
{Gomes da Silva}, J., {Santos}, N.~C., {Adibekyan}, V., {et~al.} 2021, \aap,
  646, A77, \dodoi{10.1051/0004-6361/202039765}

\bibitem[{{Guerrero} {et~al.}(2021){Guerrero}, {Seager}, {Huang}, {Vanderburg},
  {Garcia Soto}, {Mireles}, {Hesse}, {Fong}, {Glidden}, {Shporer}, {Latham},
  {Collins}, {Quinn}, {Burt}, {Dragomir}, {Crossfield}, {Vanderspek},
  {Fausnaugh}, {Burke}, {Ricker}, {Daylan}, {Essack}, {G{\"u}nther}, {Osborn},
  {Pepper}, {Rowden}, {Sha}, {Villanueva}, {Yahalomi}, {Yu}, {Ballard},
  {Batalha}, {Berardo}, {Chontos}, {Dittmann}, {Esquerdo}, {Mikal-Evans},
  {Jayaraman}, {Krishnamurthy}, {Louie}, {Mehrle}, {Niraula}, {Rackham},
  {Rodriguez}, {Rowden}, {Sousa-Silva}, {Watanabe}, {Wong}, {Zhan},
  {Zivanovic}, {Christiansen}, {Ciardi}, {Swain}, {Lund}, {Mullally},
  {Fleming}, {Rodriguez}, {Boyd}, {Quintana}, {Barclay}, {Col{\'o}n},
  {Rinehart}, {Schlieder}, {Clampin}, {Jenkins}, {Twicken}, {Caldwell},
  {Coughlin}, {Henze}, {Lissauer}, {Morris}, {Rose}, {Smith}, {Tenenbaum},
  {Ting}, {Wohler}, {Bakos}, {Bean}, {Berta-Thompson}, {Bieryla}, {Bouma},
  {Buchhave}, {Butler}, {Charbonneau}, {Doty}, {Ge}, {Holman}, {Howard},
  {Kaltenegger}, {Kane}, {Kjeldsen}, {Kreidberg}, {Lin}, {Minsky}, {Narita},
  {Paegert}, {P{\'a}l}, {Palle}, {Sasselov}, {Spencer}, {Sozzetti}, {Stassun},
  {Torres}, {Udry}, \& {Winn}}]{guerrero2021tess}
{Guerrero}, N.~M., {Seager}, S., {Huang}, C.~X., {et~al.} 2021, \apjs, 254, 39,
  \dodoi{10.3847/1538-4365/abefe1}

\bibitem[{Harris {et~al.}(2020)Harris, Millman, van~der Walt, Gommers,
  Virtanen, Cournapeau, Wieser, Taylor, Berg, Smith,
  {et~al.}}]{harris2020array}
Harris, C.~R., Millman, K.~J., van~der Walt, S.~J., {et~al.} 2020, Nature, 585,
  357

\bibitem[{{Hirsch} {et~al.}(2021){Hirsch}, {Rosenthal}, {Fulton}, {Howard},
  {Ciardi}, {Marcy}, {Nielsen}, {Petigura}, {de Rosa}, {Isaacson}, {Weiss},
  {Sinukoff}, \& {Macintosh}}]{hirsch2021understanding}
{Hirsch}, L.~A., {Rosenthal}, L., {Fulton}, B.~J., {et~al.} 2021, \aj, 161,
  134, \dodoi{10.3847/1538-3881/abd639}

\bibitem[{Huber(2017)}]{huber2017isoclassify}
Huber, D. 2017, Zenodo, doi, 10

\bibitem[{Huber {et~al.}(2017)Huber, Zinn, Bojsen-Hansen, Pinsonneault,
  Sahlholdt, Serenelli, Aguirre, Stassun, Stello, Tayar,
  {et~al.}}]{huber2017asteroseismology}
Huber, D., Zinn, J., Bojsen-Hansen, M., {et~al.} 2017, \apj, 844, 102

\bibitem[{Hunter(2007)}]{hunter2007matplotlib}
Hunter, J.~D. 2007, Computing in science \& engineering, 9, 90

\bibitem[{{Kamiaka} {et~al.}(2018){Kamiaka}, {Benomar}, \&
  {Suto}}]{kamiaka2018reliability}
{Kamiaka}, S., {Benomar}, O., \& {Suto}, Y. 2018, \mnras, 479, 391,
  \dodoi{10.1093/mnras/sty1358}

\bibitem[{Knutson {et~al.}(2014)Knutson, Fulton, Montet, Kao, Ngo, Howard,
  Crepp, Hinkley, Bakos, Batygin, Johnson, Morton, \&
  Muirhead}]{knutson_friends_2014}
Knutson, H.~A., Fulton, B.~J., Montet, B.~T., {et~al.} 2014, The Astrophysical
  Journal, 785, 126, \dodoi{10.1088/0004-637X/785/2/126}

\bibitem[{{Kraft}(1967)}]{kraft1967break}
{Kraft}, R.~P. 1967, \apj, 150, 551, \dodoi{10.1086/149359}

\bibitem[{Kunimoto \& Matthews(2020)}]{kunimoto_searching_2020}
Kunimoto, M., \& Matthews, J.~M. 2020, The Astronomical Journal, 159, 248,
  \dodoi{10.3847/1538-3881/ab88b0}

\bibitem[{Lester {et~al.}(2023)Lester, Howell, Matson, Furlan, Gnilka,
  Littlefield, Ciardi, Everett, Fajardo-Acosta, \& Clark}]{lester_visual_2023}
Lester, K.~V., Howell, S.~B., Matson, R.~A., {et~al.} 2023, The Astronomical
  Journal, 166, 166, \dodoi{10.3847/1538-3881/acf563}

\bibitem[{Mallonn {et~al.}(2022)Mallonn, Poppenhaeger, Granzer, Weber, \&
  Strassmeier}]{mallonn_detection_2022}
Mallonn, M., Poppenhaeger, K., Granzer, T., Weber, M., \& Strassmeier, K.~G.
  2022, Astronomy \& Astrophysics, 657, A102,
  \dodoi{10.1051/0004-6361/202140599}

\bibitem[{McKinney(2010)}]{mckinney2010data}
McKinney, W. 2010, in Proceedings of the 9th Python in Science Conference, Vol.
  445, Austin, TX, 51--56

\bibitem[{{Moe} \& {Kratter}(2021)}]{moe2021impact}
{Moe}, M., \& {Kratter}, K.~M. 2021, \mnras, 507, 3593,
  \dodoi{10.1093/mnras/stab2328}

\bibitem[{{Naoz}(2016)}]{Naoz2016}
{Naoz}, S. 2016, \araa, 54, 441, \dodoi{10.1146/annurev-astro-081915-023315}

\bibitem[{{Naoz} {et~al.}(2012){Naoz}, {Farr}, \& {Rasio}}]{naoz2012on}
{Naoz}, S., {Farr}, W.~M., \& {Rasio}, F.~A. 2012, \apjl, 754, L36,
  \dodoi{10.1088/2041-8205/754/2/L36}

\bibitem[{{NASA Exoplanet Archive}(2024)}]{PS_Table}
{NASA Exoplanet Archive}. 2024, Planetary Systems, Version: 2024-11-15,
  NExScI-Caltech/IPAC, \dodoi{10.26133/NEA12}

\bibitem[{{Nielsen} {et~al.}(2013){Nielsen}, {Gizon}, {Schunker}, \&
  {Karoff}}]{nielsen2013rotation}
{Nielsen}, M.~B., {Gizon}, L., {Schunker}, H., \& {Karoff}, C. 2013, \aap, 557,
  L10, \dodoi{10.1051/0004-6361/201321912}

\bibitem[{Noyes {et~al.}(1984)Noyes, Hartmann, Baliunas, Duncan, \&
  Vaughan}]{noyes_rotation_1984}
Noyes, R.~W., Hartmann, L.~W., Baliunas, S.~L., Duncan, D.~K., \& Vaughan,
  A.~H. 1984, The Astrophysical Journal, 279, 763, \dodoi{10.1086/161945}

\bibitem[{{Offner} {et~al.}(2023){Offner}, {Moe}, {Kratter}, {Sadavoy},
  {Jensen}, \& {Tobin}}]{offner2023_ppvii}
{Offner}, S.~S.~R., {Moe}, M., {Kratter}, K.~M., {et~al.} 2023, in Astronomical
  Society of the Pacific Conference Series, Vol. 534, Protostars and Planets
  VII, ed. S.~{Inutsuka}, Y.~{Aikawa}, T.~{Muto}, K.~{Tomida}, \& M.~{Tamura},
  275, \dodoi{10.48550/arXiv.2203.10066}

\bibitem[{Oliphant(2006)}]{oliphant2006guide}
Oliphant, T.~E. 2006, A guide to NumPy, Vol.~1 (Trelgol Publishing USA)

\bibitem[{Parviainen {et~al.}(2023)Parviainen, Luque, \&
  Palle}]{parviainen_spright_2023}
Parviainen, H., Luque, R., \& Palle, E. 2023, Spright: a probabilistic
  mass-density-radius relation for small planets,  arXiv,
  \dodoi{10.48550/arXiv.2311.07255}

\bibitem[{Pearce {et~al.}(2020)Pearce, Kraus, Dupuy, Mann, Newton, Tofflemire,
  \& Vanderburg}]{pearce2020orbital}
Pearce, L.~A., Kraus, A.~L., Dupuy, T.~J., {et~al.} 2020, \aj, 894, 115

\bibitem[{Queloz {et~al.}(2010)Queloz, Anderson, Collier~Cameron, Gillon, Hebb,
  Hellier, Maxted, Pepe, Pollacco, Ségransan, Smalley, Triaud, Udry, \&
  West}]{queloz_wasp-8b_2010}
Queloz, D., Anderson, D.~R., Collier~Cameron, A., {et~al.} 2010, Astronomy and
  Astrophysics, 517, L1, \dodoi{10.1051/0004-6361/201014768}

\bibitem[{{Rice} {et~al.}(2024){Rice}, {Gerbig}, \&
  {Vanderburg}}]{rice_orbital_2024}
{Rice}, M., {Gerbig}, K., \& {Vanderburg}, A. 2024, \aj, 167, 126,
  \dodoi{10.3847/1538-3881/ad1bed}

\bibitem[{{Rice} {et~al.}(2023){Rice}, {Wang}, {Gerbig}, {Wang}, {Dai},
  {Tyler}, {Isaacson}, \& {Howard}}]{rice2023orbital}
{Rice}, M., {Wang}, S., {Gerbig}, K., {et~al.} 2023, \aj, 165, 65,
  \dodoi{10.3847/1538-3881/aca88e}

\bibitem[{{Ricker} {et~al.}(2015){Ricker}, {Winn}, {Vanderspek}, {Latham},
  {Bakos}, {Bean}, {Berta-Thompson}, {Brown}, {Buchhave}, {Butler}, {Butler},
  {Chaplin}, {Charbonneau}, {Christensen-Dalsgaard}, {Clampin}, {Deming},
  {Doty}, {De Lee}, {Dressing}, {Dunham}, {Endl}, {Fressin}, {Ge}, {Henning},
  {Holman}, {Howard}, {Ida}, {Jenkins}, {Jernigan}, {Johnson}, {Kaltenegger},
  {Kawai}, {Kjeldsen}, {Laughlin}, {Levine}, {Lin}, {Lissauer}, {MacQueen},
  {Marcy}, {McCullough}, {Morton}, {Narita}, {Paegert}, {Palle}, {Pepe},
  {Pepper}, {Quirrenbach}, {Rinehart}, {Sasselov}, {Sato}, {Seager},
  {Sozzetti}, {Stassun}, {Sullivan}, {Szentgyorgyi}, {Torres}, {Udry}, \&
  {Villasenor}}]{ricker2015tess}
{Ricker}, G.~R., {Winn}, J.~N., {Vanderspek}, R., {et~al.} 2015, Journal of
  Astronomical Telescopes, Instruments, and Systems, 1, 014003,
  \dodoi{10.1117/1.JATIS.1.1.014003}

\bibitem[{{Su} \& {Lai}(2024)}]{su2024stellar}
{Su}, Y., \& {Lai}, D. 2024, arXiv e-prints, arXiv:2411.08094,
  \dodoi{10.48550/arXiv.2411.08094}

\bibitem[{Tange(2023)}]{tange_2024_11247979}
Tange, O. 2023, GNU Parallel 20240522 ('Tbilisi'),  Zenodo,
  \dodoi{10.5281/zenodo.11247979}

\bibitem[{Tokovinin \& Kiyaeva(2015)}]{tokovinin2015eccentricity}
Tokovinin, A., \& Kiyaeva, O. 2015, \mnras, 456, 2070

\bibitem[{Virtanen {et~al.}(2020)Virtanen, Gommers, Oliphant, Haberland, Reddy,
  Cournapeau, Burovski, Peterson, Weckesser, Bright,
  {et~al.}}]{virtanen2020scipy}
Virtanen, P., Gommers, R., Oliphant, T.~E., {et~al.} 2020, Nature Methods, 17,
  261

\bibitem[{Walt {et~al.}(2011)Walt, Colbert, \& Varoquaux}]{walt2011numpy}
Walt, S. v.~d., Colbert, S.~C., \& Varoquaux, G. 2011, Computing in Science \&
  Engineering, 13, 22

\bibitem[{{Wittenmyer} {et~al.}(2020){Wittenmyer}, {Wang}, {Horner}, {Butler},
  {Tinney}, {Carter}, {Wright}, {Jones}, {Bailey}, {O'Toole}, \&
  {Johns}}]{wittenmyer2020cold}
{Wittenmyer}, R.~A., {Wang}, S., {Horner}, J., {et~al.} 2020, \mnras, 492, 377,
  \dodoi{10.1093/mnras/stz3436}

\bibitem[{{Wright} {et~al.}(2012){Wright}, {Marcy}, {Howard}, {Johnson},
  {Morton}, \& {Fischer}}]{wright2012frequency}
{Wright}, J.~T., {Marcy}, G.~W., {Howard}, A.~W., {et~al.} 2012, \apj, 753,
  160, \dodoi{10.1088/0004-637X/753/2/160}

\bibitem[{{Zanazzi} \& {Lai}(2018)}]{Zanazzi2018torquedamp}
{Zanazzi}, J.~J., \& {Lai}, D. 2018, \mnras, 477, 5207,
  \dodoi{10.1093/mnras/sty951}

\bibitem[{{Zhang} {et~al.}(2023){Zhang}, {Weiss}, {Huber}, {Jensen}, {Brandt},
  {Collins}, {Conti}, {Isaacson}, {Lewin}, {Marino}, {Massey}, {Murgas},
  {Palle}, {Radford}, {Relles}, {Srdoc}, {Stockdale}, {Tan}, \&
  {Wang}}]{Zhang2023dynamical}
{Zhang}, J., {Weiss}, L.~M., {Huber}, D., {et~al.} 2023, arXiv e-prints,
  arXiv:2310.03299, \dodoi{10.48550/arXiv.2310.03299}

\bibitem[{{Zhang} {et~al.}(2018){Zhang}, {Li}, {Xie}, {Zhou}, {Liu}, \&
  {Zhang}}]{Zhang2018}
{Zhang}, Y., {Li}, Q., {Xie}, J.-W., {et~al.} 2018, \apj, 861, 116,
  \dodoi{10.3847/1538-4357/aac6c3}

\end{thebibliography}
\bibliographystyle{aasjournal}

\begin{longrotatetable}
    \begin{deluxetable*}{rrrrrrrrrrr}
        \tablecaption{Systems in our edge-on binaries sample, ordered in reverse by angular separation between the two stars. Listed parameters include the \textit{Gaia} DR3 source IDs of the stellar components, along with the coordinates of the primary star ($\rm{RA_p}$,$\rm{Dec_p}$), the angular and sky-projected separation between the binary components ($\theta$ and $s$, respectively), the \textit{Gaia} $G$ magnitudes ($G_\mathrm{p}$, $G_\mathrm{s}$) and effective temperatures ($T_\mathrm{eff,p}$, $T_\mathrm{eff,s}$) of both stars, and the derived binary inclination ($i$). Additional columns can be found in the full table.\label{chartable}}
        \tabletypesize{\scriptsize}
        \tablehead{
            Prim. Gaia DR3 ID & Sec. Gaia DR3 ID & $\text{RA}_\mathrm{p}$ ($^{\circ}$) & $\text{Dec}_\mathrm{p}$ ($^{\circ}$) & $G_\mathrm{p}$ & $G_\mathrm{s}$ & $s$ (au) & $\theta$ (\arcsec) & $T_\mathrm{eff,p}$ (K) &  $T_\mathrm{eff,s}$ (K) & $i$ ($^{\circ}$) \\
        }
        \startdata
        3127503620545728000 & 3127574019351694848 & $100.54683$ & $3.58014$  & $10.9$ & $12.0$ & $759$ & $50.05$ & $3598\pm91$ & $3398\pm129$ & $92.9\substack {+4.1 \\ -0.9}$\\
6603693881832177664 & 6603693808817829888 & $341.24247$ & $-33.25103$  & $10.7$ & $11.8$ & $747$ & $35.82$ & $3747\pm89$ & $3561\pm98$ & $87.2\substack {+0.9 \\ -3.7}$\\
3053492881541641728 & 3053492881541639680 & $113.02360$ & $-8.88207$  & $5.8$ & $9.9$ & $683$ & $23.81$ & $6109\pm597$ & $4187\pm137$ & $90.1\substack {+0.2 \\ -0.2}$\\
3341002562175024640 & 3340814889285222400 & $83.34474$ & $12.35416$  & $11.6$ & $13.6$ & $584$ & $20.18$ & $3711\pm88$ & $3375\pm127$ & $94.3\substack {+6.5 \\ -1.8}$\\
4503423641091792896 & 4503423641091795968 & $268.93686$ & $18.50017$  & $8.7$ & $10.8$ & $416$ & $18.50$ & $4460\pm171$ & $3764\pm91$ & $88.0\substack {+1.2 \\ -4.3}$\\
6840822904599512064 & 6840822904602986496 & $328.28073$ & $-12.82863$  & $10.3$ & $11.9$ & $417$ & $17.30$ & $3934\pm106$ & $3593\pm92$ & $93.6\substack {+6.0 \\ -1.0}$\\
3042300093686461440 & 3042299715729340928 & $118.35642$ & $-8.65503$  & $11.3$ & $11.6$ & $763$ & $16.72$ & $4041\pm111$ & $3958\pm107$ & $88.8\substack {+0.6 \\ -2.6}$\\
6020602863959507968 & 6020602868279750656 & $248.30340$ & $-35.15661$  & $12.3$ & $12.7$ & $764$ & $14.74$ & $3873\pm104$ & $3781\pm95$ & $94.5\substack {+7.4 \\ -1.5}$\\
3619215425323808256 & 3619215425323808768 & $209.64778$ & $-8.16940$  & $13.7$ & $13.8$ & $718$ & $13.53$ & $3580\pm94$ & $3573\pm96$ & $89.9\substack {+0.9 \\ -1.1}$\\
4522817514378193408 & 4522817514378191872 & $273.89399$ & $16.23973$  & $11.4$ & $12.7$ & $437$ & $13.13$ & $3816\pm94$ & $3589\pm93$ & $92.0\substack {+4.2 \\ -0.8}$\\
4314903404672032768 & 4314903198513595904 & $292.41089$ & $11.30184$  & $12.9$ & $13.5$ & $462$ & $12.67$ & $3584\pm94$ & $3486\pm116$ & $92.6\substack {+3.4 \\ -1.4}$\\
1827242816201846272 & 1827242816176111360 & $300.18212$ & $22.70974$  & $7.4$ & $13.2$ & $226$ & $11.44$ & $5021\pm202$ & $3330\pm111$ & $88.2\substack {+0.6 \\ -1.8}$\\
1859483069783082496 & 1859483069781443072 & $312.35957$ & $30.69763$  & $10.1$ & $13.6$ & $570$ & $10.92$ & $4739\pm189$ & $3598\pm92$ & $87.7\substack {+0.9 \\ -4.4}$\\
2966316109264052224 & 2966316109264051200 & $88.62609$ & $-19.70445$  & $7.3$ & $10.1$ & $250$ & $10.73$ & $5260\pm223$ & $3984\pm109$ & $92.9\substack {+8.2 \\ -0.9}$\\
4604269610637580800 & 4604269606340411392 & $268.95853$ & $33.41959$  & $9.7$ & $12.3$ & $315$ & $10.61$ & $4281\pm154$ & $3600\pm91$ & $93.3\substack {+9.9 \\ -1.3}$\\
6498800193370215424 & 6498800197665444864 & $350.91573$ & $-56.58772$  & $12.1$ & $12.2$ & $727$ & $10.50$ & $4096\pm115$ & $4069\pm114$ & $91.7\substack {+3.5 \\ -0.5}$\\
4375470410650198016 & 4375470414949031936 & $264.98106$ & $1.18925$  & $12.1$ & $13.1$ & $741$ & $9.90$ & $4144\pm111$ & $3839\pm96$ & $87.6\substack {+1.2 \\ -6.7}$\\
2070269864129830400 & 2070269864129829888 & $309.65164$ & $44.64113$  & $9.5$ & $10.2$ & $532$ & $9.83$ & $5147\pm222$ & $4744\pm196$ & $87.4\substack {+0.7 \\ -4.6}$\\
3482326708703712768 & 3482326708703712768 & $173.06825$ & $-29.26038$  & $5.5$ & $5.6$ & $269$ & $9.58$ & $6220\pm754$ & $6162\pm672$ & $91.0\substack {+3.2 \\ -0.4}$\\
4305953826646567936 & 4305953826646567424 & $286.44880$ & $6.54648$  & $6.8$ & $8.7$ & $591$ & $9.41$ & $6737\pm1251$ & $5745\pm301$ & $89.1\substack {+0.5 \\ -2.3}$\\
1270385943872155136 & 1270386008294438400 & $229.33094$ & $25.86152$  & $12.7$ & $12.9$ & $758$ & $9.41$ & $3990\pm109$ & $3950\pm107$ & $86.9\substack {+1.2 \\ -6.5}$\\
3623497335919494144 & 3623497331624720384 & $199.36348$ & $-10.54622$  & $6.8$ & $11.8$ & $708$ & $9.25$ & $6923\pm1318$ & $4280\pm154$ & $91.5\substack {+3.6 \\ -0.8}$\\
5913288643608847360 & 5913288643608845312 & $259.64377$ & $-62.60066$  & $9.3$ & $11.3$ & $672$ & $9.25$ & $5554\pm243$ & $4474\pm174$ & $86.5\substack {+1.1 \\ -5.5}$\\
6610154852675341312 & 6610154852675341312 & $344.45165$ & $-26.10948$  & $7.5$ & $9.0$ & $590$ & $9.18$ & $6147\pm650$ & $5521\pm238$ & $89.8\substack {+0.3 \\ -0.6}$\\
5554798134403542016 & 5554798134403542016 & $95.34041$ & $-46.73626$  & $10.0$ & $12.7$ & $753$ & $9.17$ & $5313\pm228$ & $4029\pm111$ & $91.7\substack {+3.4 \\ -0.9}$\\
3871910272860536832 & 3871910272859979264 & $164.34765$ & $12.60530$  & $13.2$ & $13.4$ & $454$ & $9.12$ & $3635\pm87$ & $3607\pm90$ & $91.5\substack {+2.7 \\ -0.6}$\\
6356247449972639744 & 6356247449972640768 & $320.28866$ & $-77.17484$  & $13.5$ & $13.6$ & $438$ & $8.81$ & $3597\pm92$ & $3579\pm95$ & $89.8\substack {+0.2 \\ -0.4}$\\
5596377845938835456 & 5596002538831140864 & $121.71961$ & $-30.76897$  & $12.1$ & $12.7$ & $444$ & $8.64$ & $3876\pm100$ & $3744\pm89$ & $88.5\substack {+0.7 \\ -3.8}$\\
840526822110467968 & 840526753390990976 & $177.67738$ & $53.47455$  & $9.5$ & $13.4$ & $589$ & $8.42$ & $5395\pm229$ & $3731\pm89$ & $94.3\substack {+9.8 \\ -2.0}$\\
1973090800219117056 & 1973090044304874240 & $326.37226$ & $44.23037$  & $12.7$ & $13.8$ & $667$ & $8.32$ & $4005\pm111$ & $3728\pm89$ & $89.5\substack {+0.4 \\ -1.4}$\\

        \enddata
        \label{tab:binary_systems}
        \tablecomments{See an online version of this manuscript for a downloadable version of the full table.}
    \end{deluxetable*}
\end{longrotatetable}



\end{document}